\documentclass[journal]{IEEEtran}
\usepackage{cite}
\usepackage{amsmath}
\usepackage{algorithmic}
\usepackage{graphicx}
\usepackage{textcomp}
\usepackage{csquotes}
\usepackage{xcolor,soul}
\usepackage{adjustbox}
\usepackage{hyperref}
\usepackage{multirow}
\usepackage{titlesec}
\usepackage{subcaption}
\usepackage{makecell}

\usepackage[numbers,sort&compress]{natbib}

\usepackage{etoolbox}
\usepackage{adjustbox}

\definecolor{columbiablue}{rgb}{0.61, 0.87, 1.0}
\titlespacing{\subsubsection}{0pt}{0pt}{1pt}

\usepackage{todonotes}
\usepackage{url}
\usepackage{algorithmic}
\usepackage{algorithm}
\usepackage{gensymb}
\usepackage{bm}

\newcounter{algsubstate}

\newlength\myindent
\begin{document}

\title{Multiple and Gyro-Free Inertial Datasets}

\author{Zeev~Yampolsky\IEEEauthorrefmark{1},
        Yair~Stolero,
        Nitzan~Pri-Hadash,
        Dan~Solodar,
        Shira~Massas,
        Itai~Savin,
        and~Itzik~Klein~
\thanks{All authors are with the Hatter Department of Marine Technologies, Charney School of Marine Sciences, University of Haifa.\\
\IEEEauthorrefmark{1}Corresponding author: Z. Yampolsky (email: zyampols@campus.haifa.ac.il).}}


\maketitle
\begin{abstract}
 An inertial navigation system (INS) utilizes three orthogonal accelerometers and gyroscopes to determine platform position, velocity, and orientation. There are countless applications for INS, including robotics, autonomous platforms, and the internet of things. Recent research explores the integration of data-driven methods with INS, highlighting significant innovations, improving accuracy and efficiency. Despite the growing interest in this field and the availability of INS datasets, no datasets are available for gyro-free INS (GFINS) and multiple inertial measurement unit (MIMU) architectures. To fill this gap and to stimulate further research in this field, we designed and recorded GFINS and MIMU datasets using 54 inertial sensors grouped in nine inertial measurement units. These sensors can be used to define and evaluate different types of MIMU and GFINS architectures. The inertial sensors were arranged in three different sensor configurations and mounted on a mobile robot and a passenger car. In total, the dataset contains $35$ hours of inertial data and corresponding ground truth trajectories. The data and code are freely accessible through our GitHub repository.
\end{abstract}

\section{Introduction}\label{background_intro_sec}
The inertial navigation system (INS) is a navigation technology which relies on measuring the specific force and angular velocity vectors of a platform to determine its position, velocity, and orientation. Typically three accelerometers and three gyroscopes, arranged in orthogonal triads, form the inertial measurement unit (IMU). In recent years, micro-electro-mechanical systems (MEMS) IMUs have gained popularity due to their compact size, low cost, and low power consumption. However, the accuracy of a MEMS INS degrades rapidly over time due to sensor errors and accumulated uncertainties. According to current technology, these errors are more pronounced in MEMS-based gyroscopes than in MEMS-based accelerometers.
\newline 
Over the past few years, there has been a surge of interest in exploring the potential of multiple IMU (MIMU) systems, which consist of arrays of interconnected IMUs. Unlike a single IMU, MIMU are composed of multiple sensors which are tightly integrated and aligned. These IMUs, when combined using sophisticated data fusion algorithms, exhibit a collective output which surpasses those of individual units. A comprehensive exploration of this burgeoning field is provided in a literature review made by Nilsson and Skog \cite{nilsson2016inertial}. 
By harnessing data fusion algorithms on MIMU outputs, researchers aim to achieve three primary objectives: 1) enhancing detection accuracy, 2) minimizing errors, particularly IMU sensor noise and 3) redundancy in case of a sensor failure.
The effectiveness of MIMUs extends across various domains, including coarse alignment \cite{larey2020multiple}, calibration tasks \cite{rehder2016extending, carlsson2021self}, integration with global navigation satellite
systems (GNSS) \cite{luciani2022mimu} \cite{guerrier2009improving}, pedestrian navigation \cite{skog2014open,bancroft2010multiple, skog2014pedestrian, bose2017noise}, data fusion, filtering operations \cite{patel2022multi, patel2021sensor, bancroft2011data, libero2022unified}, and localization algorithms \cite{zhang2020lightweight}.
\newline Another method to mitigate the performance issues associated with MEMS gyroscopes is the adoption of a gyro-free system, known as GFINS \cite{pachter2013gyro, schuler1967measuring, liu2014design, mostov2000design, zhou2021gyro}.
In this configuration, the conventional trio of orthogonal gyroscopes gives way to the utilization of $N$ linear accelerometers. These accelerometers play a dual role, computing both specific force and angular acceleration vectors, effectively transforming the system into a functional INS. 
The effectiveness of a GFINS relies on the quality, quantity, and arrangement of the accelerometers \cite{zappa2001number, tan2001design}. Notice, however that to accurately estimate the motion and orientation of a rigid body, a minimum of six distributed uni-axial units is required \cite{chen1994gyroscope}. In GFINS, another integration (on the calculated angular acceleration vector) is added to the measured angular acceleration, increasing sensor noise (and other noise terms) impacting the system's accuracy significantly.
\newline An emerging field in navigation and specifically in inertial navigation focuses on the use of machine learning and deep learning algorithms for enhancing the system accuracy and robustness. While conventional model-based approaches may yield satisfactory results when applied to high-end inertial sensors, learning-based approaches open the door to accurate navigation using low-cost commercial sensors and for longer periods of time. Research in data-driven inertial sensing can be categorized by the platform type used, which highly affects the inertial sensor error model and accuracy. For example, Tlio \cite{liu2020tlio} is a deep neural network for regressing 3D position displacement and its uncertainty for pedestrian navigation while Vertzberger et al. \cite{vertzberger2021attitude} provides an attitude estimation for pedestrians. LLIO \cite{wang2022llio} is a lightweight version of Tlio, aimed for real-time usage on low power end devices. QuadNet \cite{shurin2022quadnet} is a deep-learning approach based on convolution neural network (CNN), which estimates the position of a quad rotor maneuvering in a periodic motion, using only low-cost inertial sensors. VIO-DualProNet \cite{solodar2023vio} is an adaptive learning-based technique designed to estimate inertial process noise covariance of a drone and improve visual-inertial odometry fusion accordingly. Brossard et al. \cite{brossard2020denoising} presented a deep learning framework aimed at denoising gyroscope readings for open-loop attitude estimation for drones. Buchanan et al. \cite{buchanan2022deep} proposed a learning-based method for estimating IMU bias factors in pedestrian and quadruped robot visual-inertial factor graph problems. BeamsNet \cite{cohen2022beamsnet} is a data-driven approach for estimating an autonomous underwater vehicle (AUV) velocity using high-end IMU and DVL measurements as inputs. DeepLIO \cite{iwaszczuk2021deeplio} is a learning-based method for LiDAR-inertial odomtery of ground robots. Brossard et al. \cite{brossard2019learning} used a deep neural network for enhancing wheel odometer and IMU fusion of wheeled robots.
\newline To research and apply learning-based methods in the inertial navigation domain large datasets are required. Learning-based methods rely on the ability of the models to learn patterns and characteristics in the data, making the need for good datasets critical. These datasets should cover diverse scenarios, capturing various environmental conditions, motion dynamics, and sensor setups. A well-designed dataset helps researchers thoroughly train models and validate their performance, improve their methods and enhance navigation performance. Inertial datasets can be found in various domains for aerial, and maritime platforms.
Existing publicly available datasets includes the RONIN dataset~\cite{herath2020ronin}  containing more than 40 hours of IMU sensor data from 100 human subjects, with ground-truth 3D trajectories under natural human motion. The RIDI dataset~\cite{yan2018ridi} is a wide and versatile inertial dataset for pedestrian navigation. The authors of the OXIOD dataset~\cite{chen2018oxiod} collected 158 sequences of handheld inertial data, covering a distance of over 42 km. The inertial ADVIO dataset~\cite{cortes2018advio} contains 23 separate recordings captured in six different locations
with a total length of 4.47 kilometers, and a total duration of 1
hour and 8 minutes. The Kitti dataset \cite{geiger2013vision} is a popular benchmark dataset in autonomous vehicles navigation, containing 6 hours of traffic scenarios with a sensor suite including inertial sensors, cameras, and GNSS. The IO-VNBD dataset \cite{onyekpe2021io} is a diverse inertial dataset collected using a car, with more than 58 hours and 4,400 traveled kilometers of data.
A popular dataset for inertial navigation research for drones is the Euroc dataset \cite{burri2016euroc}, consisting of drone recordings in various motions and scenarios. NTU-viral \cite{nguyen2022ntu} is a versatile aerial platform dataset, consisting of both indoor and outdoor recordings of various sensors including IMUs, cameras, and ranging sensors. The autonomous platforms inertial dataset \cite{shurin2022autonomous} includes data recordings from multiple platform types and contains 805.5 minutes of data from quadrotors, underwater vehicles, land vehicles, and boats.
\newline While the above described datasets are wide and cover different types of platforms and sensor types, they are all aimed for methods using a single IMU and aren't applicable to MIMU and GFINS architectures. MIMU methods require multiple inertial sensors working simultaneously while GFINS requires at least six accelerometers arranged in a specific configuration. Hence, there is a need for an inertial dataset aimed at MIMU and GFINS architectures to allow application of data-driven approaches and to prosper research in the field. To fill this gap, this paper presents the multiple and gyro free - inertial dataset (MAGF-ID), a large and versatile dataset for MIMU and GFINS research. The MAGF-ID consists of inertial recordings made with nine IMUs, totalling 54 inertial sensors, arranged in three different configurations. These configurations were mounted on a land vehicle and a mobile robot recording 115 trajectories in different dynamics. The dataset contains 35 hours of inertial recordings and associated ground-truth trajectories. 
To the best of our knowledge, MAGF-ID is the first available dataset aiming for GFINS and MIMU research.
\newline The rest of the paper is organized as follows. Section \ref{scien_sec} describes the scientific background and mathematical formulation of the INS equations of motion and the MIMU and GFINS architectures. Section \ref{methods_section} discusses the measurement sensors, MAGF-ID configurations, experimental platforms, and the recording protocol. Next, Section \ref{data_valid_sec} visualizes the collected data and presents an overview of the collected data. Finally, Section \ref{data_Recs} provides the description of the dataset, its structure, and its files.

\section{Scientific Background}\label{scien_sec}
This section aims to present and describe the mathematical formulation of the MIMU and GFINS equations of motion (EoM). To that end, we first introduce the INS EoM.
\subsection{INS Equations of Motion}\label{scien_sec_ins_eof}
The INS equations of motion calculate the navigation state from the inertial sensor readings. The navigation state is comprised of the position vector $\textbf{\textit{p}}^l$ expressed in the local coordinate frame, the velocity vector $\textbf{\textit{v}}^l$ expressed in the local coordinate frame and transformation matrix $\textbf{T}_{b}^l$ between the body frame and the local frame. The change of the position vector is given by\cite{titterton2004strapdown, farrell1999global}:
\begin{equation}\label{eq:ins_p}
\dot{\textbf{\textit{p}}}^l = \textbf{\textit{v}}^l
\end{equation}
The rate of change of the velocity vector is denoted by:
\begin{equation}\label{eq:ins_v}
\dot{\textbf{\textit{v}}}^l = \textbf{T}_{b}^l \textbf{\textit{f}}^b + \textbf{\textit{g}}^l
\end{equation}
where $g^l$ is the local gravity vector. Notice that in \eqref{eq:ins_v} the earth rotation rate and transport rate are neglected as low-performance inertial sensors are addressed. The rate of change of the transformation matrix is:
\begin{equation}\label{eq:ins_t}
\dot{\textbf{T}}_{b}^{l} = \textbf{T}_{b}^{l}\boldsymbol{{\Omega}}_{ib}^{b}
\end{equation}
where $\mathbf{{\Omega}}_{ib}^b$ is the skew-matrix of the angular velocity vector:
\begin{equation}
\bm{\Omega}_{ib}^b = \left[ \begin{array}{ccc}
0 & -\omega_{z} & \omega_{y} \\
\omega_{z} & 0 & -\omega_{x} \\
-\omega_{y} & \omega_{x} & 0 \\
\end{array} \right]
\end{equation}
and $\omega_{x} , \omega_{y}, \omega_{z}$ are the angular velocity components in the $x,y,z$ directions.
\subsection{MIMU Equations of Motion}\label{scien_sec_mimu}
Typical MIMU architecture consists of $\textit{N}$ IMUs sampled simultaneously. Under the assumption that all IMUs are relatively close to each other, an average measurement is made to reduce the measurement noise:
\begin{equation}\label{eq:mimu_f}
\textbf{\textit{f}}_{\text{MIMU}} = \frac{1}{N}\sum_{i=1}^{N}\textbf{\textit{f}}_i
\end{equation}
\begin{equation}\label{eq:mimu_omega}
\bm{\omega}_{\text{MIMU}} = \frac{1}{N}\sum_{i=1}^{N}\bm{\omega}_i
\end{equation}
where $N$ is the number of IMUs in the MIMU configuration. Next, the average inertial measurement from equations \eqref{eq:mimu_f}-\eqref{eq:mimu_omega}
are plugged into the INS EoM  \eqref{eq:ins_v}-\eqref{eq:ins_t} to derive the MIMU navigation solution. 

\subsection{GFINS Equations of Motion}\label{scien_sec_gfins}
The GFINS calculates the specific force and angular acceleration vectors through the accelerometers' measurements.
The specific force measurement vector of a GFINS consisting of $N$ accelerometers is given by:
\begin{equation}
\centering
\textbf{Y} = [\textbf{\textit{f}}_1, \textbf{\textit{f}}_2, \ldots, \textbf{\textit{f}}_N]^T
\end{equation}
where $\textbf{\textit{f}}_i$ is the specific force measured by the $i\textsuperscript{th}$ accelerometer. The configuration matrix $\textbf{H}$ of a GFINS can be written as:
\begin{equation}
\textbf{H} = \begin{bmatrix}
(\textbf{\textit{i}}_1 \times \textbf{\textit{e}}_1)^T & \textbf{\textit{e}}_1^T \\
\vdots & \vdots \\
(\textbf{\textit{i}}_N \times \textbf{\textit{e}}_N)^T & \textbf{\textit{e}}_N^T
\end{bmatrix}_{Nx6}
\end{equation}
where $\textbf{\textit{i}}_i$ is the lever-arm of the $i\textsuperscript{th}$ accelerometer relative to the GFINS output point and $\textbf{\textit{e}}_i$ is the direction of the $i\textsuperscript{th}$ accelerometer. 
\newline The GFINS output is an estimate of the specific force and angular acceleration vectors:
\begin{equation}\label{eq:gfins_output}
\begin{bmatrix}
\dot{\bm{\omega}} \\
\textbf{\textit{f}}
\end{bmatrix}
= \begin{bmatrix}
\textbf{H}_{\dot{\omega}} \\
\textbf{H}_a
\end{bmatrix}(\textbf{Y} - \textbf{M})
\end{equation}
where $\dot{\bm{\omega}}$ is the angular acceleration. The matrix $\textbf{M}$ relates between the body angular velocity and the GFINS model:
\begin{equation}
\textbf{M} = \begin{bmatrix}
{\textbf{\textit{i}}_1}^T\bm{\Omega}_{ib}^b{\textbf{\textit{e}}_1} \\
\vdots \\
{\textbf{\textit{i}}_N}^T\bm{\Omega}_{ib}^b{\textbf{\textit{e}}_N}
\end{bmatrix}
\end{equation}
To form the GFINS EoM, the estimated specific force vector, \eqref{eq:gfins_output}, is plugged into \eqref{eq:ins_p}. As the EoM requires the angular velocity in \eqref{eq:ins_v}, the estimated angular acceleration from \eqref{eq:gfins_output} is integrated. Thus, an additional integration is added to the INS equations.

\section{Proposed Approach}\label{methods_section}
This section describes the making of the MAGF-ID dataset in detail. The Subsection \ref{sensors_subsec} describes the sensors used in all of the recordings along with their specifications. Subsection \ref{config_subsec} presents a detailed description of the three MAGF-ID configurations, labeled  \textbf{C1-C3}. The  Subsection \ref{platforms_subsec} presents the two platforms used in our experiments. Finally, the Subsection \ref{rec_protocol_sec} describes the employed protocol in each trajectory recording.\\

\subsection{Sensor Types}\label{sensors_subsec}
To conduct our experiments, we employed nine Xsens DOT IMUs \cite{Xsens_dots}, to obtain inertial raw measurements, while relying on the Inertial Labs MRU-P \cite{Inertial_labs_MRU} for ground truth (GT). As noted, the MRU-P is equipped with a licensed GNSS Real-Time Kinematic (RTK) TerraStar-C Pro system \cite{TerraStar_C_Pro}. We validated its accuracy through a static experiment, achieving a positioning accuracy of less than $30$ cm while sampling at $5$Hz. The specifications of the MRU-P and Xsens DOT inertial sensors are provided in Table \ref{table_sensor_comp}.
\begin{table}[h!]
\caption{Error terms description of the used inertial sensors in the data collection experiment.}\label{table_sensor_comp}
\begin{adjustbox}{width=1.0\columnwidth,center}\begin{tabular}{|c|c|c|c|c|c|}
\hline
    & \begin{tabular}[c]{@{}c@{}}MRU-P\\ Accelerometer\end{tabular} & \begin{tabular}[c]{@{}c@{}}MRU-P\\ Gyroscope\end{tabular} & \begin{tabular}[c]{@{}c@{}}MRU-P\\ Magnetometer\end{tabular} & \begin{tabular}[c]{@{}c@{}}Xsens DOT\\ Accelerometer\end{tabular} & \begin{tabular}[c]{@{}c@{}}Xsens DOT\\ Gyroscope\end{tabular} \\ \hline
\begin{tabular}[c]{@{}c@{}}Sampling\\ Rate {[}Hz{]}\end{tabular} & 100     & 100  & 100    & 120     & 120     \\ \hline
\begin{tabular}[c]{@{}c@{}}Bias in-run\\ Stability\end{tabular}  & $0.005$ ${[}mg{]}$   & $1$ ${[}\degree/hour{]}$  & $0.2$ ${[}\mu T{]}$  & $0.03 {[}mg{]}$    & $10 {[}\degree/hour{]}$        \\ \hline
Noise Density   & $0.025 {[}mg/\sqrt{Hz}{]} $ &$ 0.004 {[}\degree/s/\sqrt{Hz}{]}$  & $0.3 {[}\mu T/\sqrt{Hz}{]} $  & $120 {[}mg/\sqrt{Hz}{]}$ & $0.007 {[}\degree/s/\sqrt{Hz}{]}$  \\ \hline
\end{tabular}
\end{adjustbox}
\end{table}

\subsection{Sensor Configurations}\label{config_subsec} 
\begin{figure}[h!]
    \centering    
    \includegraphics[width=0.7\linewidth, clip, keepaspectratio]{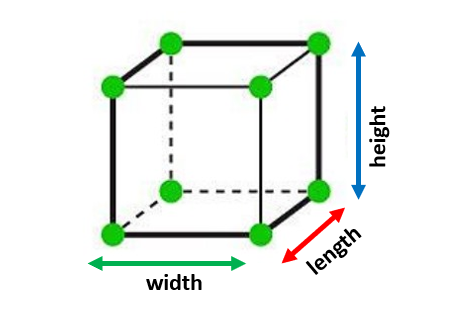}
    \caption{The box-shaped geometry is common to all three configurations, \textbf{C1-C3}. Each green dot represents a DOT sensor.}
    \label{conf_description}
\end{figure}

\begin{figure}[ht!]
    \begin{adjustbox}{minipage=\linewidth, scale=1.0,center}
        \begin{subfigure}{.49\linewidth}        \includegraphics[width=\linewidth, clip, keepaspectratio]{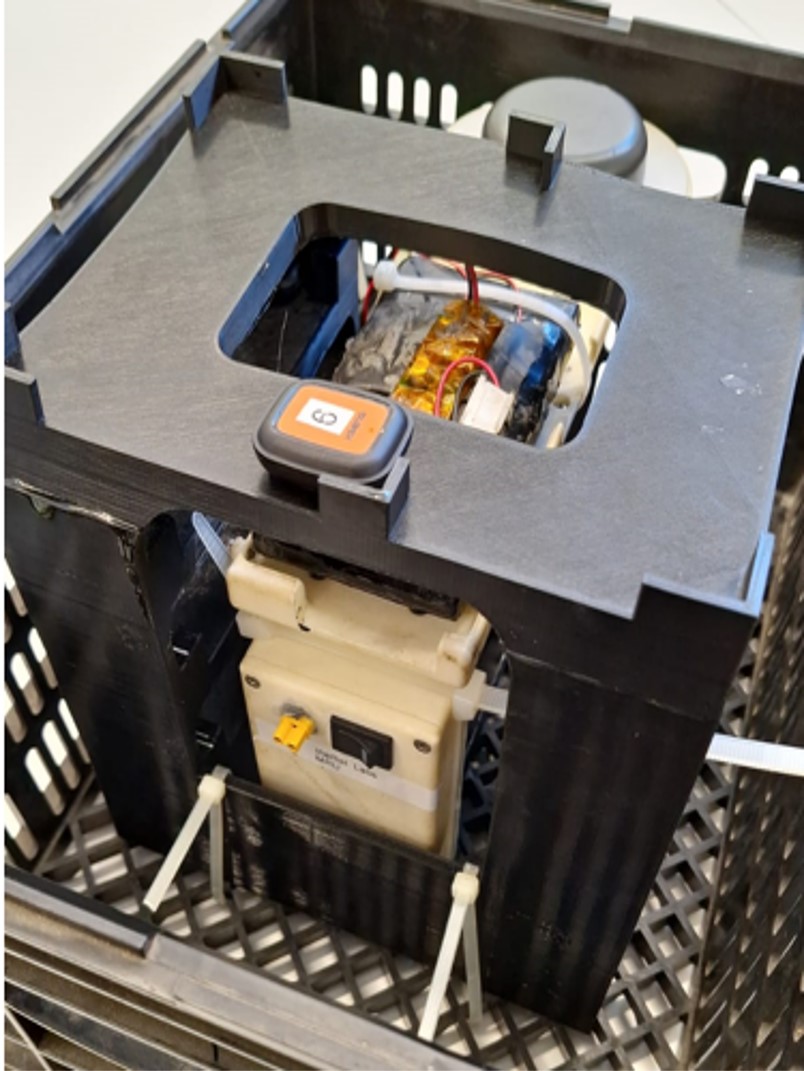}
          \caption{\textbf{C1} Top-Left View}
          \label{image_conf1_TOP}
        \end{subfigure}
        \begin{subfigure}{.49\linewidth} 
        \includegraphics[width=\linewidth, clip, keepaspectratio]{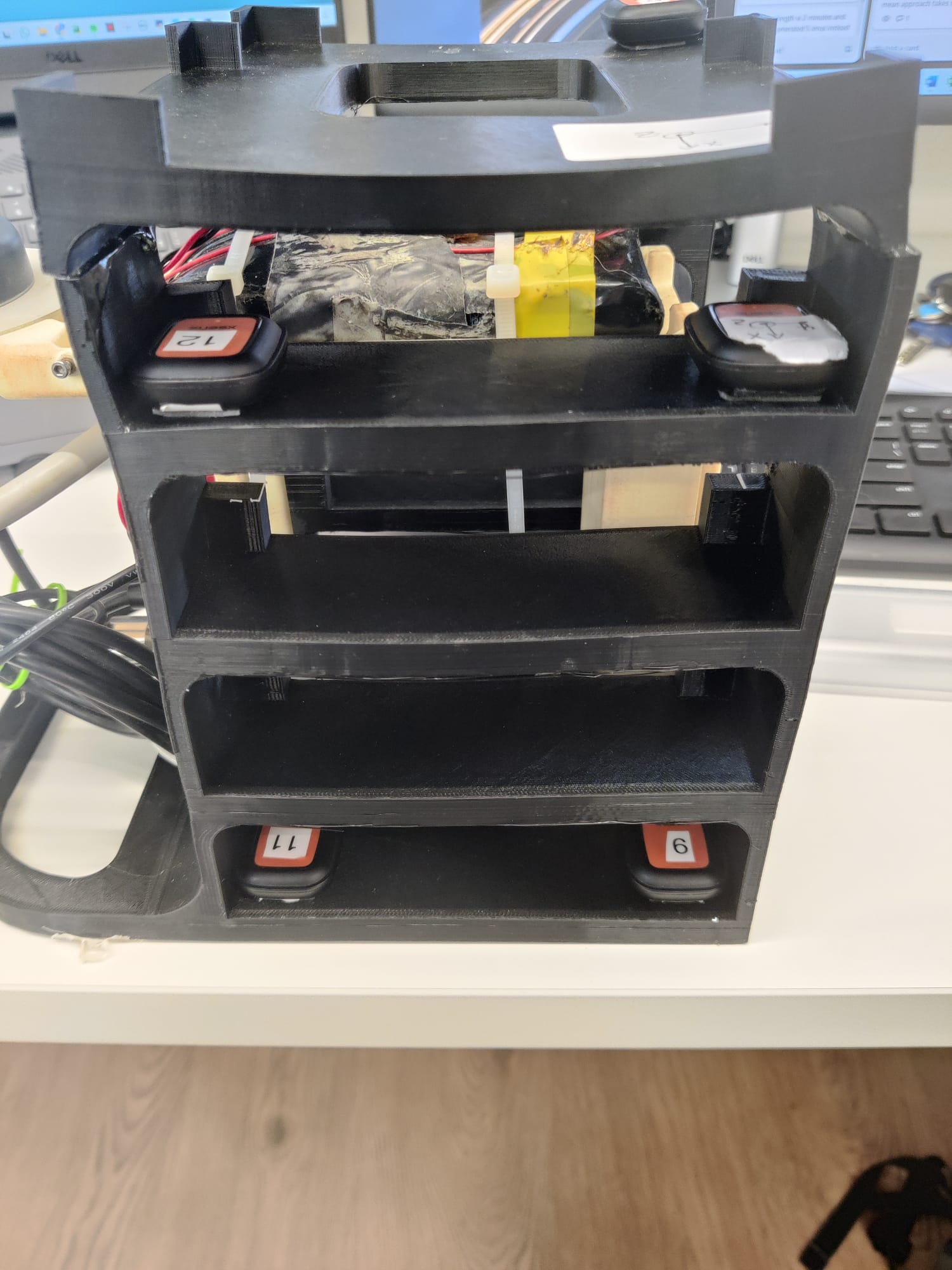}
          \caption{\textbf{C1} Front View}
        \label{image_conf1_FRONT}
        \end{subfigure}
    \caption{\ref{image_conf1_TOP} \textbf{C1} - top-left view showing the MRU-P fitting snugly inside the DOT structure, and \ref{image_conf1_FRONT} front view showing four out of the eight DOTs.}
    \label{MRU_and_DOT_conf1}
    \end{adjustbox}
\end{figure}

To solve the equations presented in Subsection \ref{scien_sec_gfins}, a minimum of six sensors are required. In order to provide redundancy, improved accuracy, and measurements at different locations, we designed three configurations consisting of eight or nine DOT IMUs, labeled \textbf{C1}, \textbf{C2}, and \textbf{C3}. In each configuration, the DOTs were positioned in the shape of a box, as illustrated in Figure \ref{conf_description}, and oriented in the same direction. The dimensions of each configuration were used to determine the distances between pairs of DOTs. In configuration \textbf{C1}, the DOTs were placed on a 3D printed structure, which was designed to fit snugly around the MRU-P. There, eight DOTs were placed in the aforementioned box configuration and a ninth DOT was added to at the top, as seen in Figure \ref{MRU_and_DOT_conf1}.
\begin{figure}[h!]
    \begin{adjustbox}{minipage=\linewidth, scale=1,center}
        \begin{subfigure}[t]{.3\textwidth}
        \includegraphics[width=0.9\linewidth, clip, keepaspectratio]{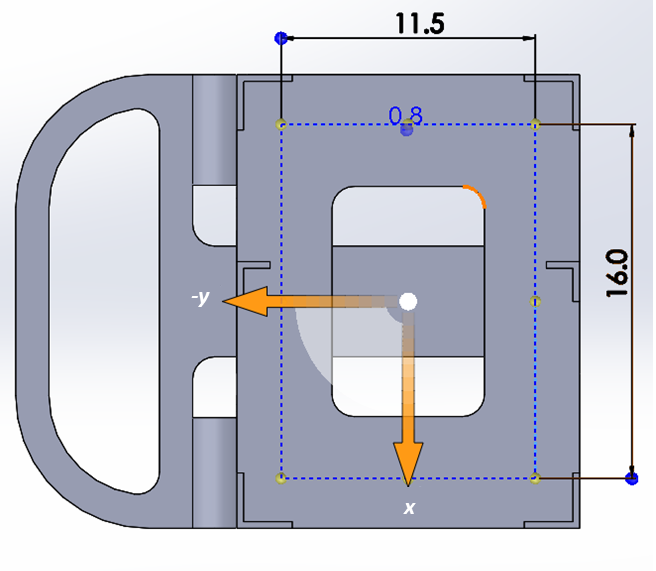}
          \caption{\textbf{C1} Top View}
          \label{conf1_TOP}
        \end{subfigure}
        \begin{subfigure}[t]{.3\textwidth} 
        \includegraphics[width=0.9\linewidth, clip, keepaspectratio]{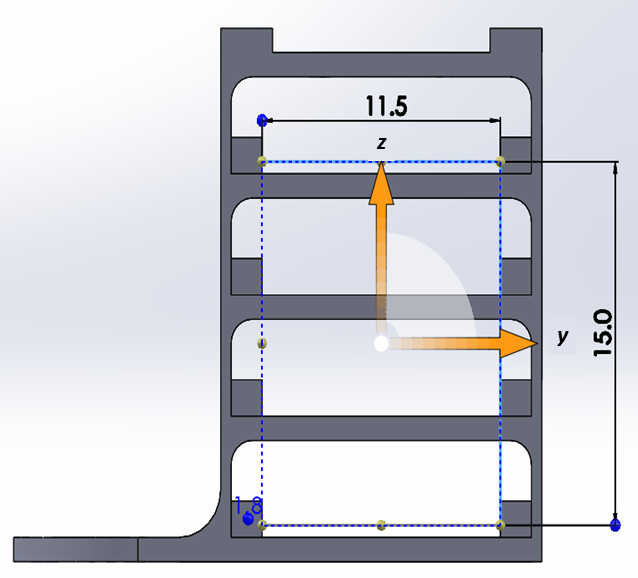}
          \caption{\textbf{C1} Front View}
          \label{conf1_FRONT}
        \end{subfigure}
        \begin{subfigure}[t]{.3\textwidth} 
        \includegraphics[width=0.9\linewidth, clip, keepaspectratio]{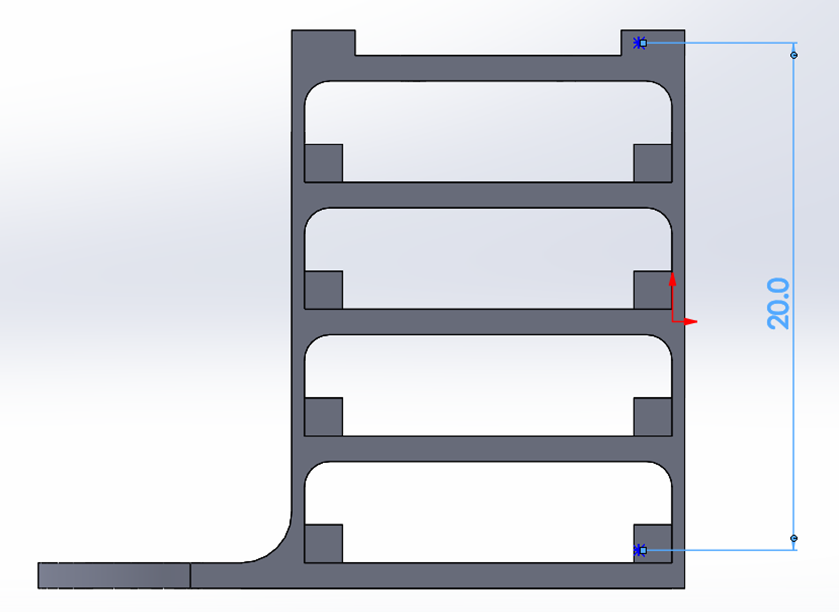}
          \caption{\textbf{C1} Front View, height of ninth DOT}
          \label{conf1_FRONT_ninthDOT}
        \end{subfigure}
    \caption{Dimensions of \textbf{C1}, sub Figure \ref{conf1_TOP} top view showing the length along the $x$-axis and the width along the $y$-axis, sub Figure \ref{conf1_FRONT} front view showing the width along the $y$-axis and the height along the $z$-axis, and sub Figure \ref{conf1_FRONT_ninthDOT} front view showing the height of the ninth DOT.}
    \label{key_dimensions_conf1}
    \end{adjustbox}
\end{figure}
\begin{figure}[h!]
    \begin{adjustbox}{minipage=\linewidth, scale=1,center}
        \begin{subfigure}[t]{.45\linewidth}
        \includegraphics[width=\linewidth, clip, keepaspectratio]{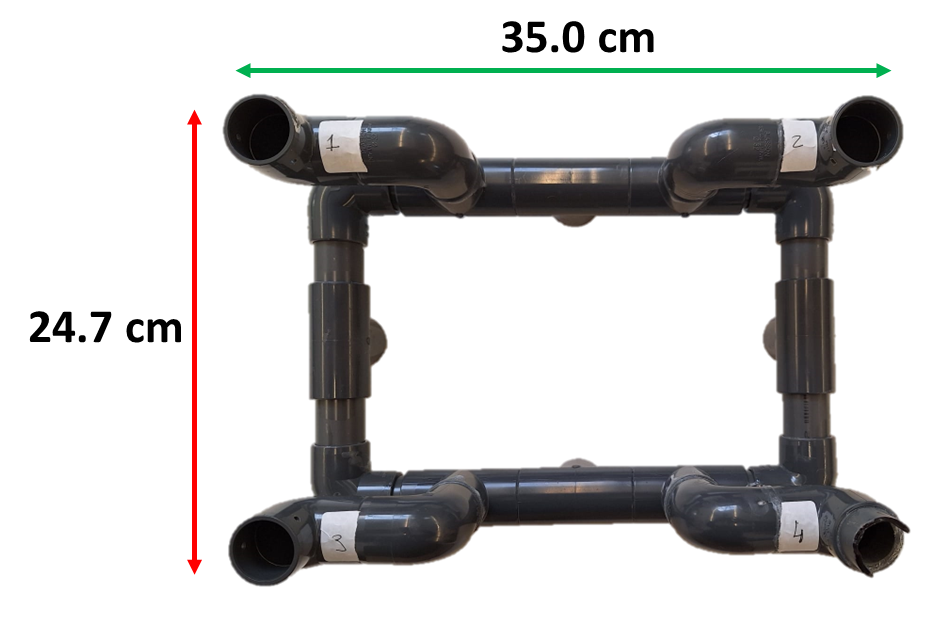}
          \caption{\textbf{C2} Top View}
          \label{conf2_TOP}
        \end{subfigure}
        \begin{subfigure}[t]{.49\linewidth} 
        \includegraphics[width=\linewidth, clip, keepaspectratio]{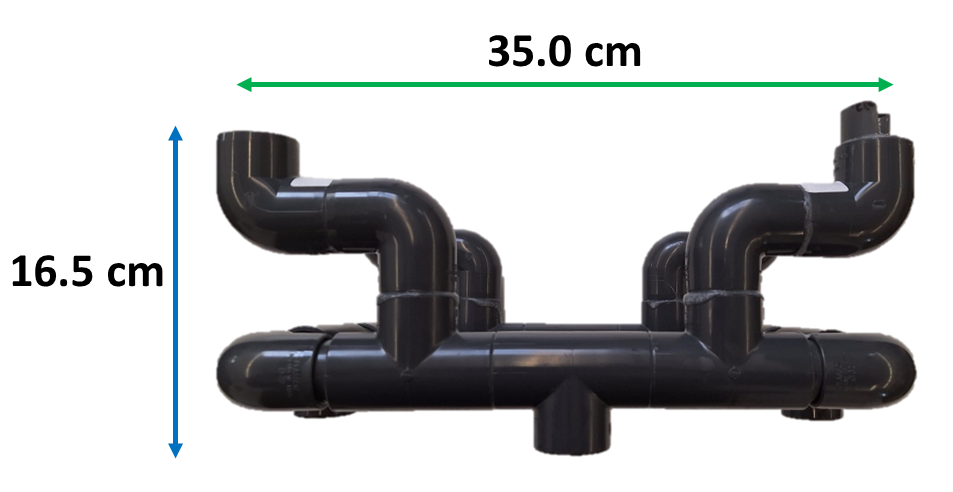}
          \caption{\textbf{C2} Front View}
          \label{conf2_FRONT}
        \end{subfigure}
        \vspace{0.1\linewidth}
    \caption{Dimensions of \textbf{C2} sub Figure \ref{conf2_TOP} top view showing the length along the $x$-axis in red, and the width along the $y$-axis in green, and sub Figure \ref{conf2_FRONT} the front view showing the width along the $y$-axis in green and the height along the $z$-axis in blue.}
    \label{key_dimensions_conf2}
    \end{adjustbox}
\end{figure}
\begin{figure}[h!]
    \begin{adjustbox}{minipage=\linewidth, scale=1,center}
        \begin{subfigure}[t]{.45\linewidth}          \includegraphics[width=\linewidth, clip, keepaspectratio]{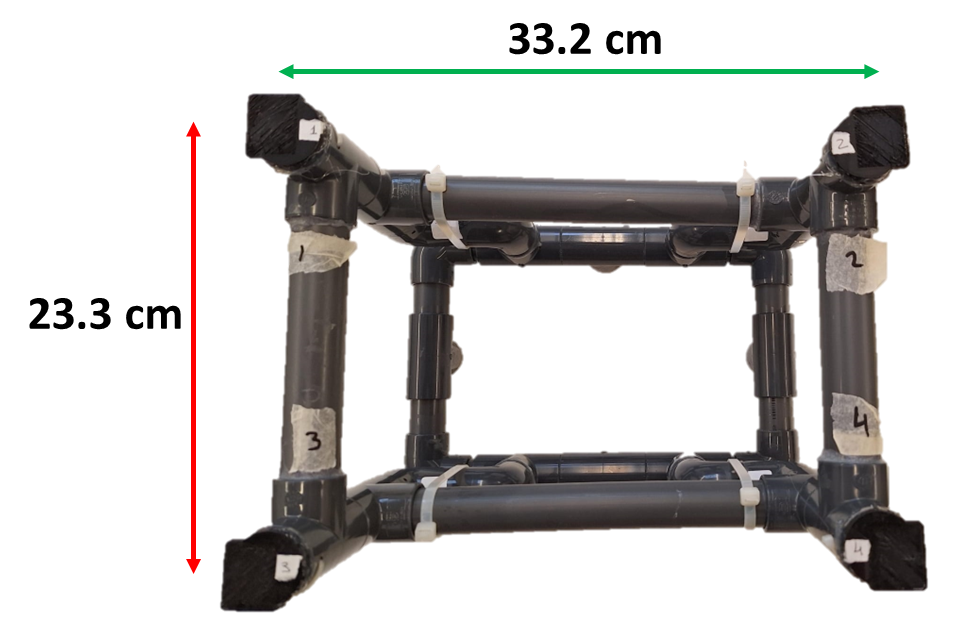}
          \caption{\textbf{C3} Top View}
          \label{conf3_TOP}
        \end{subfigure}
        \begin{subfigure}[t]{.49\linewidth}
        \includegraphics[width=\linewidth, clip, keepaspectratio]{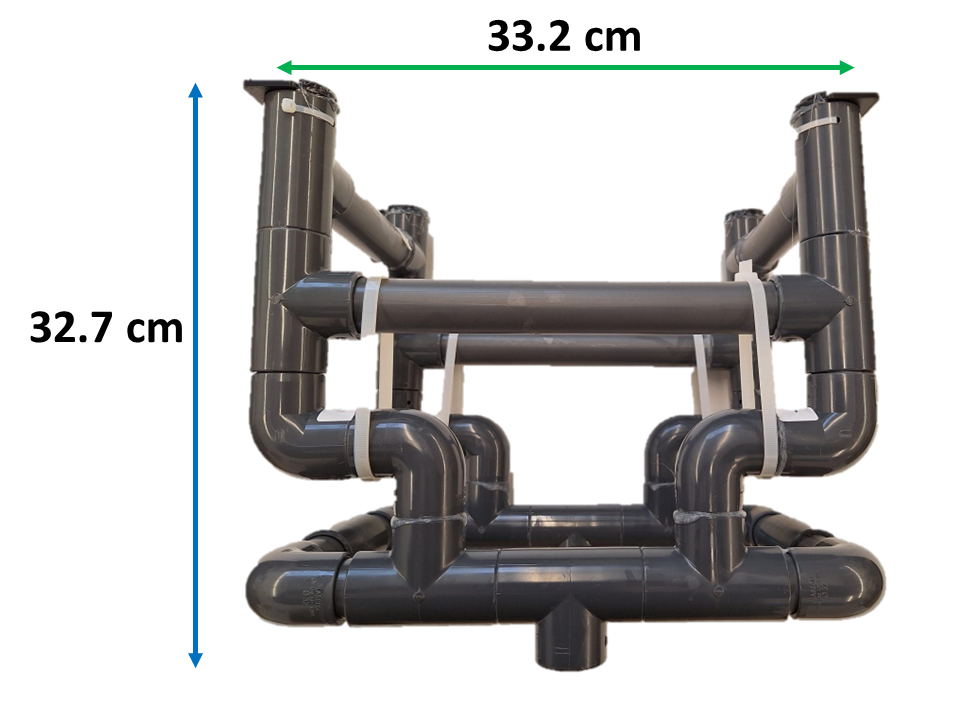}
          \caption{\textbf{C3} Front View}
          \label{conf3_FRONT}
        \end{subfigure}
    \caption{Dimensions of \textbf{C3}, sub Figure \ref{conf3_TOP} top view showing the length along the $x$-axis in red, and the width along the $y$-axis in green, and sub Figure \ref{conf3_FRONT} front view showing the width along the $y$-axis in green and the height along the $z$-axis in blue.}
    \label{key_dimensions_conf3}
    \end{adjustbox}
\end{figure}
\begin{figure}[h!]
\begin{adjustbox}{minipage=\linewidth, scale=1,center}
    \begin{subfigure}[t]{.45\linewidth}          
    \includegraphics[trim = {2cm 2cm 2cm 2cm},clip,width=\linewidth,keepaspectratio]{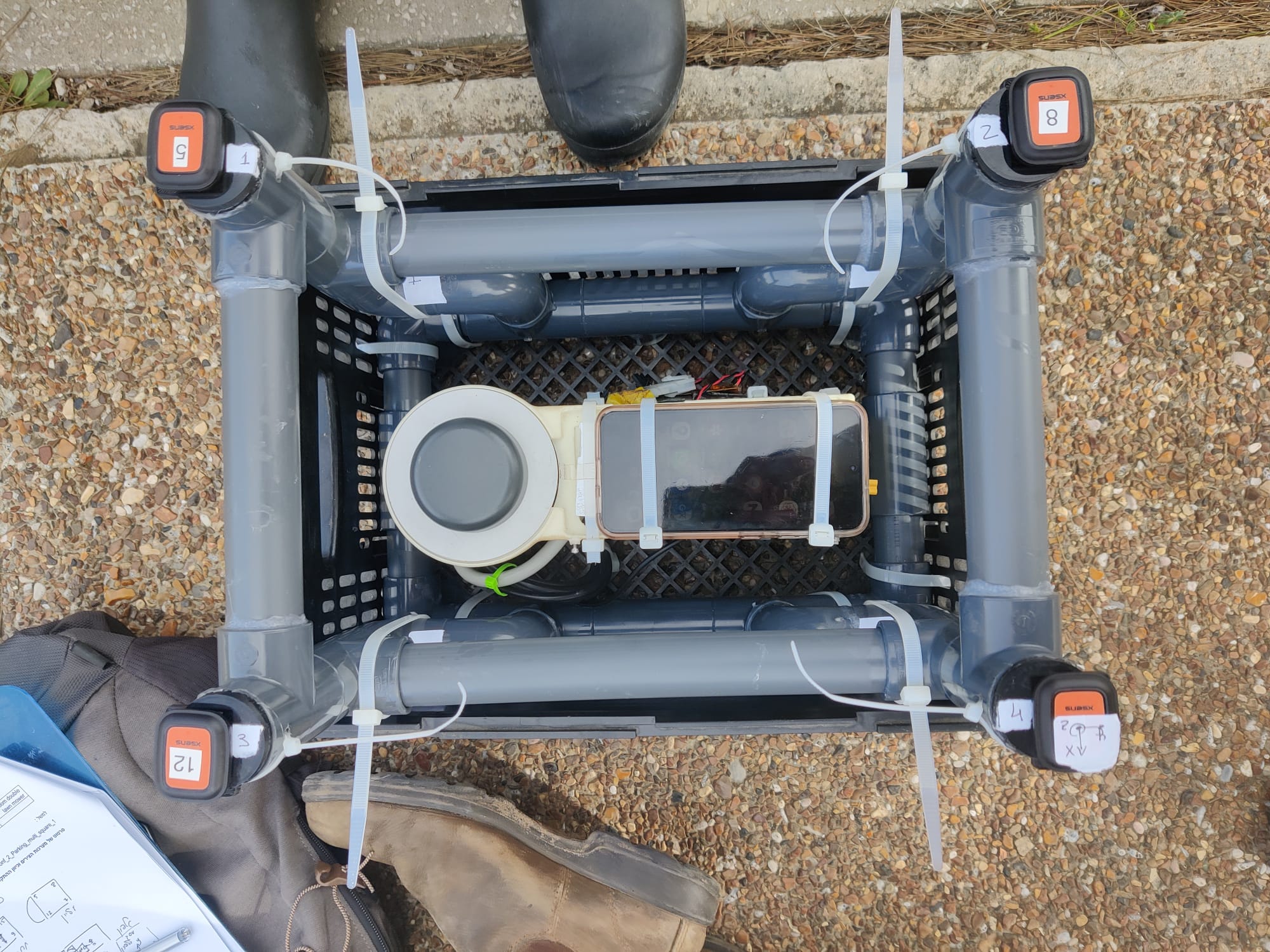}
    \caption{Placement of MRU-P and PVC structure inside crate.}
    \label{MRU_PVC_in_crate}
    \end{subfigure}
    \hspace{0.1\linewidth}
    \begin{subfigure}[t]{.49\linewidth} 
        \includegraphics[width=\linewidth, clip, keepaspectratio]{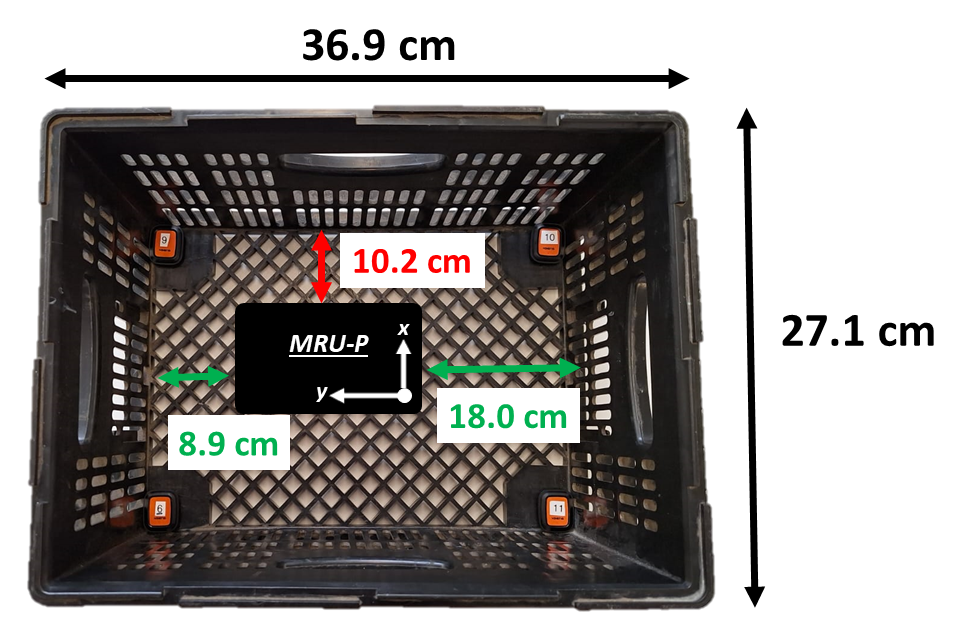}
          \caption{Position of the MRU-P inside of the crate.}
          \label{MRU_dim_in_crate}
    \end{subfigure}
    \caption{ Sub Figure \ref{MRU_PVC_in_crate} Placement of MRU-P inside of the crate when using \textbf{C3} configuration, and sub Figure \ref{MRU_dim_in_crate} shows an illustration showing the exact location of the MRU-P inside of the crate, relative to the crate.}
    \end{adjustbox}
\end{figure}
\begin{table}[h!]
    \centering
    \caption{The dimensions of configurations \textbf{C1-3} used to determine the distances between each pairs of DOTs in a configuration.}
    \label{lever_arm_dist}
    \begin{tabular}{|c || c | c | c|}
        \hline
        Configuration & Length [cm] & Width [cm] & Height [cm] \\
        \hline\hline
        \textbf{C1} & $16.0$ & $11.5$ & $15.0$ \\
        \hline
        \textbf{C2} & $24.7$ & $35.0$ & $16.5$ \\ 
        \hline
        \textbf{C3} & $23.3$ & $33.2$ & $32.7$ \\ 
        \hline
    \end{tabular}
\end{table}
\\ Front and top views of \textbf{C1} and its dimensions are presented in Figure \ref{key_dimensions_conf1}. In configuration \textbf{C2}, the spacing of the DOTs was increased in all three dimensions. Therefore, a larger PVC rig was constructed, as shown in Figure \ref{key_dimensions_conf2}. Configuration \textbf{C3} was taller than configuration \textbf{C2}, but had the same base dimensions. So, a modular PVC extension was added on top, as seen in Figure \ref{key_dimensions_conf3}. The dimensions of all three configurations can be found in Table \ref{lever_arm_dist}.
Nine DOTs were utilized in each experiment using \textbf{C1}, whereas eight DOTs were utilized in experiments using \textbf{C2} and \textbf{C3}.
The dimensions provided in Table \ref{lever_arm_dist} correspond to the box-shaped geometry, common to configurations \textbf{C1-C3} as illustrated in Figure \ref{conf_description}. The MRU-P and the structures holding the DOTs in their configurations were placed in a perforated crate and fastened with zip-ties. This created a rigid unit, as seen in Figure \ref{MRU_PVC_in_crate}. The MRU-P's exact position was measured relative to the inner walls of the crate, as seen in Figure \ref{MRU_dim_in_crate}.
\subsection{Platforms}\label{platforms_subsec}
\begin{figure}[h!]
    \begin{adjustbox}{minipage=\linewidth, scale=1}
    \hspace{0.025\linewidth}
        \begin{subfigure}[t]{0.45\linewidth}        \includegraphics[width=\linewidth, clip, keepaspectratio]{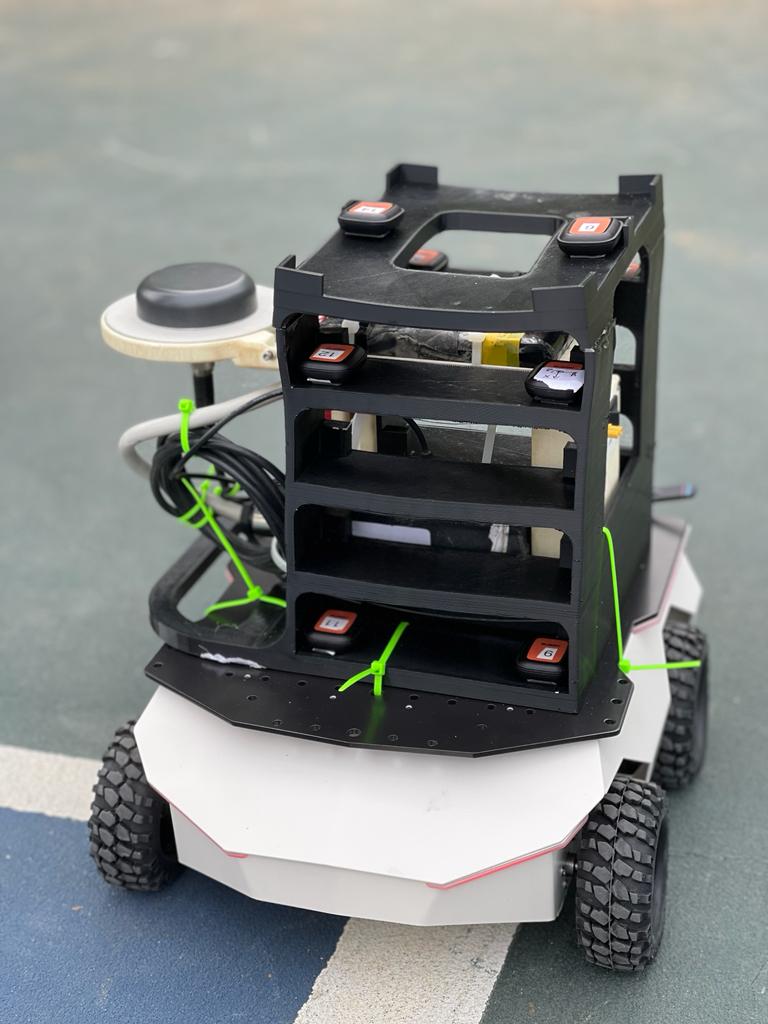}
          \caption{ROSbot Front View}
          \label{rosbot_front_view_with_conf_1}
        \end{subfigure}
        \begin{subfigure}[t]{0.45\linewidth} \includegraphics[width=\linewidth, clip, keepaspectratio]{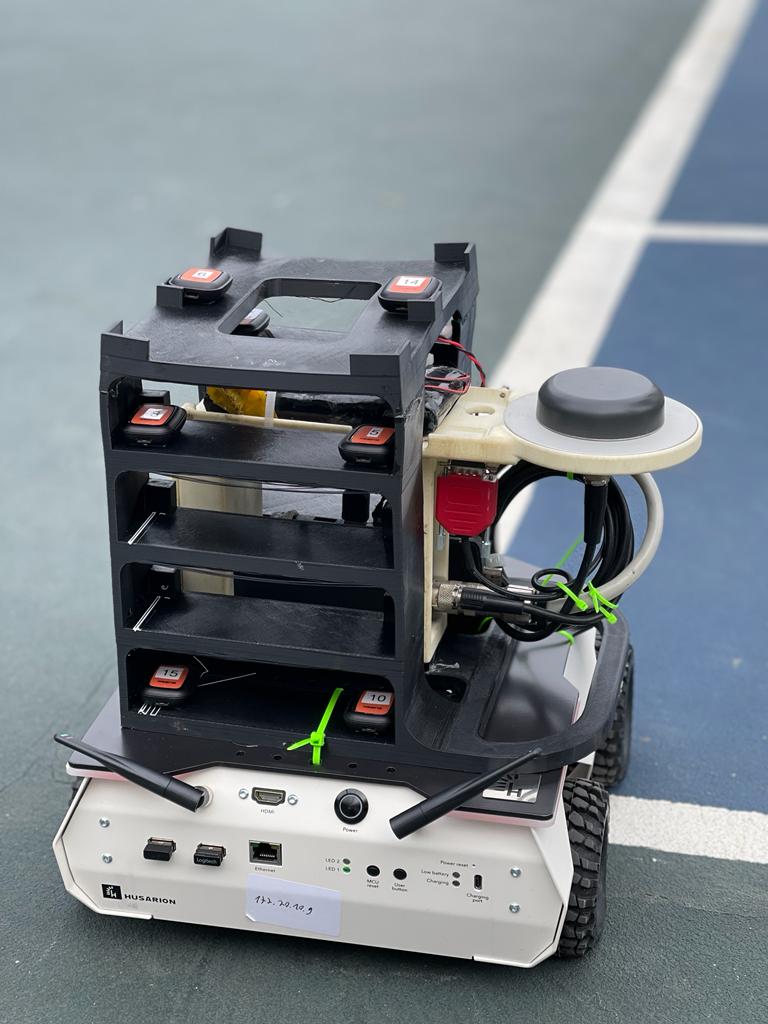}
          \caption{ROSbot Back Side View}
          \label{rosbot_back_view_with_conf_1}
        \end{subfigure}
    \caption{Mounting of \textbf{C1} configuration onto the ROSbot during an experiment.}
    \label{rosbot_with_conf_1_images}
    \end{adjustbox}
\end{figure}

Two different mobile platforms were used in this work. The first platform was used to record the dynamics of a larger vehicle operated manually, in this case, a car. The second platform was designed to mimic dynamics resembling a mobile robot operating in a small area, such as a mall or a large hall. Both platforms employed in this work were mounted with configurations \textbf{C1-C3}. The two platforms are described below:
\begin{itemize}
    \item \textbf{Land Vehicle}: A Ford Fiesta passenger car was used in the experiments. All three configurations were mounted on the roof of the car, just above the passenger seat. The configurations were fastened securely in a plastic crate, and the crate was strapped tightly to the car using two lines, each secured with a ratchet. Figure \ref{conf_1_mounting_oncar_images} shows one of our experiment setups mounted to the roof of the car. Figure \ref{car_and_configs_mounting_measurements} shows the dimensions of the car and the position of the configuration.

%
\begin{figure}[h!]
    \begin{adjustbox}{minipage=\linewidth, scale=1.0 , center}
        \begin{subfigure}[t]{.46\linewidth}        \includegraphics[width=\linewidth, clip, keepaspectratio]{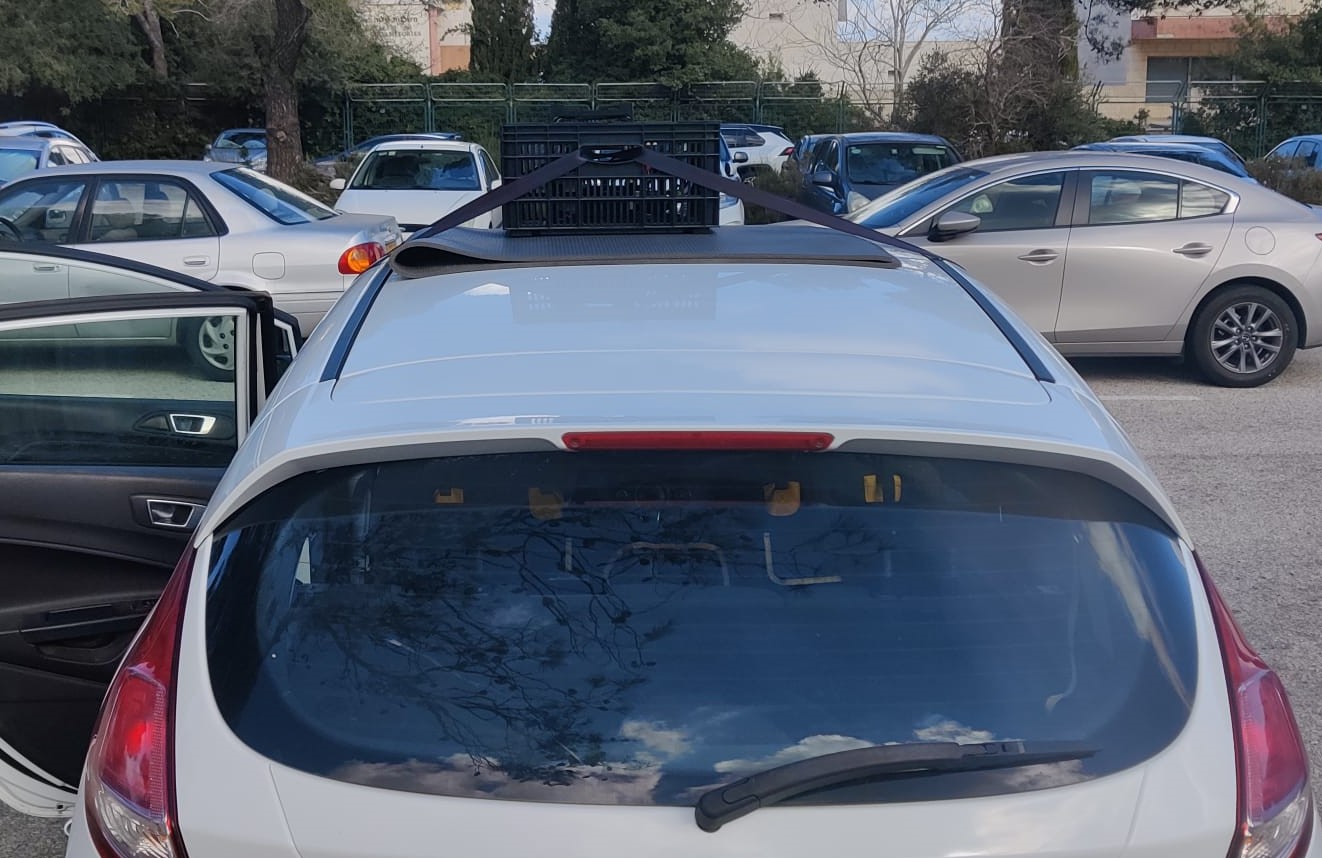}
          \caption{Back Side View}
          \label{car_back_view_gf_mount}
        \end{subfigure}
        \begin{subfigure}[t]{.53\linewidth} \includegraphics[width=\linewidth, clip, keepaspectratio]{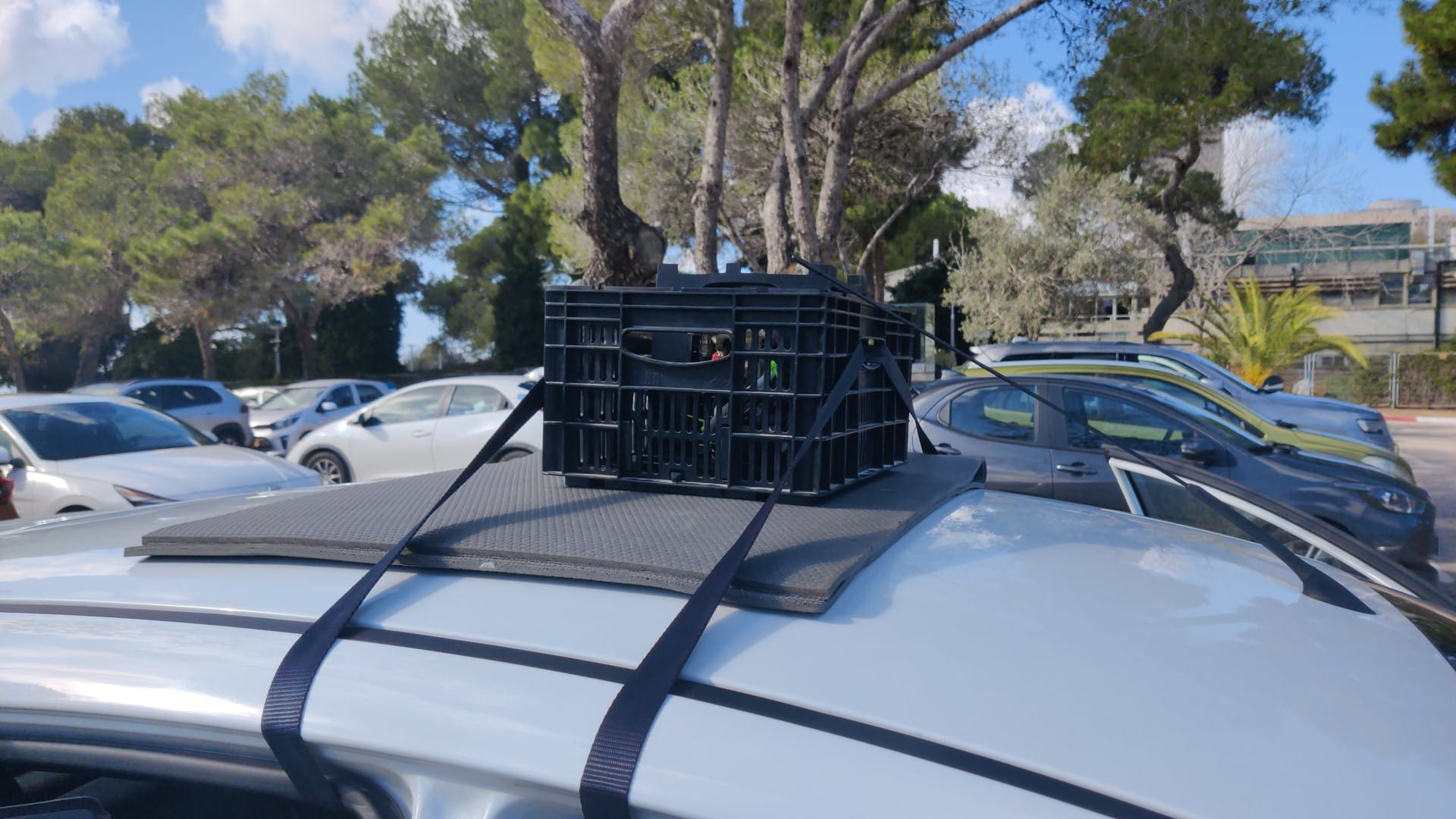}
          \caption{Left Side View}
          \label{car_left_side_view_of_the_config}
        \end{subfigure}
    \caption{Images showing the mounting of \textbf{C1} configuration om the car.}
    \label{conf_1_mounting_oncar_images}
    \end{adjustbox}
\end{figure}
\begin{figure}[h!]
    \begin{adjustbox}{minipage=\linewidth, scale=1 , center}
        \begin{subfigure}[t]{.59\linewidth}        \includegraphics[width=\linewidth, clip, keepaspectratio]{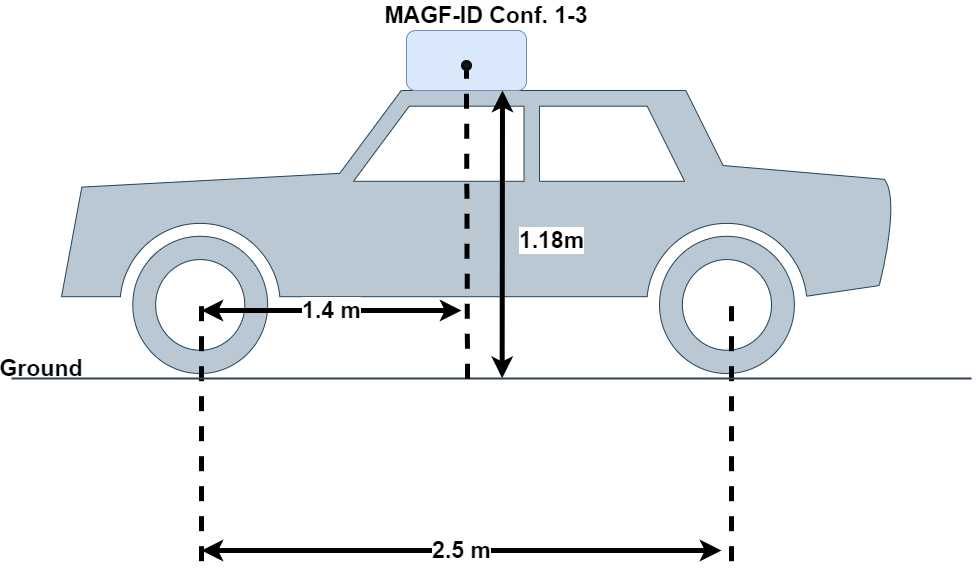}
          \caption{Car Mounting Left Side}
          \label{car_mounting_measurments_left}
        \end{subfigure}
        \begin{subfigure}[t]{.39\linewidth} \includegraphics[width=\linewidth, clip, keepaspectratio]{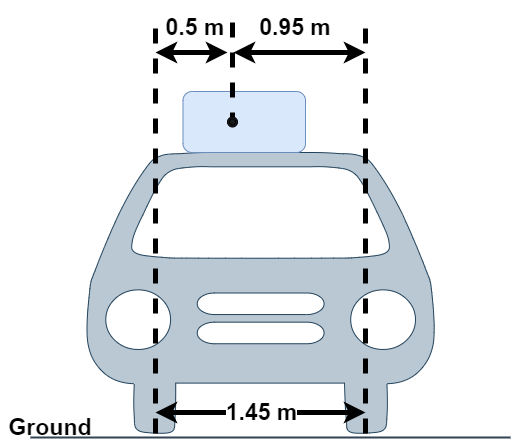}
          \caption{Car Mounting Back Side}
          \label{car_mounting_measurments_front}
        \end{subfigure}
    \caption{Illustration of the mounting position of the crate on top of the car.}
    \label{car_and_configs_mounting_measurements}
    \end{adjustbox}
\end{figure}

\item  \textbf{Mobile Robot}: The Husarion ROSbot XL robot, presented in Figure \ref{rosbot_with_conf_1_images}, was used in our experiments. It was manually remote controlled, because the sensors which enable autonomous driving had to be removed to make way for the crate with the sensors. The ROSbot XL platform has dimensions of $332 \times 325 \times 133.5$ mm [L $\times$ W $\times$ H] \cite{ROSbotXL_dim}. It can be remote controlled at three preset speeds, henceforth referred to as slow, medium and fast. 
\end{itemize}

\subsection{Recording Protocol}\label{rec_protocol_sec}
\begin{figure}
    \centering
    \includegraphics[width=0.9\linewidth, clip, keepaspectratio]{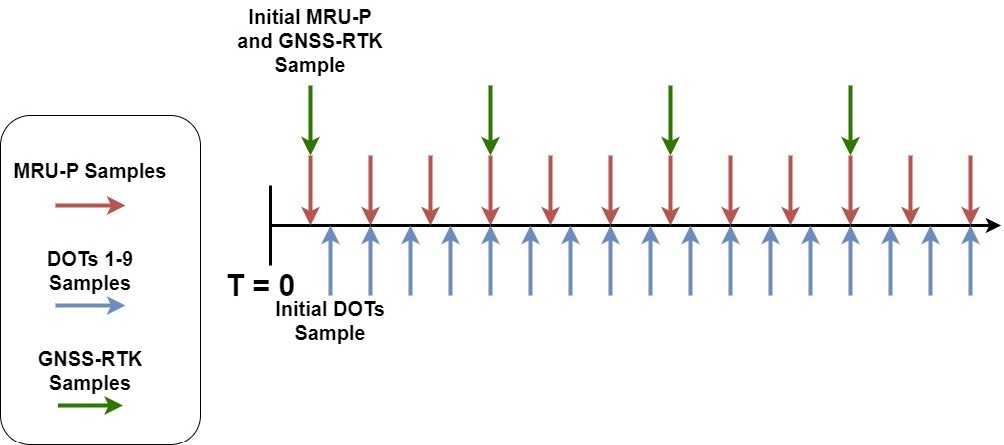}
    \caption{Illustration of the different sampling frequency rates of the DOTs and MRU-P sensors. It is also evident at the beginning that there is a short delay caused by the manual operation.}
    \label{samples_times_fig}
\end{figure}

The purpose of this section is to describe the recording protocol employed in each data collection experiment as part of the MAGF-ID dataset creation. To describe the recording protocol, the reasoning for using such a protocol will be presented first, followed by a detailed description of the protocol.\\
In each of the experiments, two  sensors were used: the DOT IMUs and the MRU-P. All of the DOTs were connected to an application which controlled and synchronized them.  Using this application allowed us to start the DOT IMUs recordings simultaneously and with the same sampling rate. An operator initiated the DOT recordings manually. As for the MRU-P inertial sensors and GNSS-RTK,  the slower sampling rate of the GNSS-RTK (5Hz) was aligned with the faster sampling rates of the inertial sensors (100Hz) at the overlapping measurements. In this case as well, the recording of the MRU-P was initiated by an operator. Thus, a slight delay is to be excepted in the recordings of the different sensors. Moreover, the different sampling rates of the DOT IMUs and the MRU-P leads to an additional challenge in ensuring that all recordings of the different sensors sample the same physical conditions at the same time.  Figure \ref{samples_times_fig} illustrated the difference in the sensors' sampling rates.
\\
To synchronize between the MRU-P and the DOT IMUs, Protocol \ref{recording_protocol} was implemented. The main idea of the protocol was to create a strong signal by shaking the configuration at the beginning of each experiment, imposing stationary conditions before and after the movement. As a result, it is possible to align the measurements between the DOTs and the MRU-P based on that signal, during the post-processing of the recordings.

\begin{algorithm}[h!]
\floatname{algorithm}{Protocol}
    \begin{algorithmic}[1]
        \STATE Start MRU-P and DOT recordings simultaneously
        \STATE Standstill for $15$ seconds
        \STATE Shake the MAGF-ID configuration 
        \STATE Standstill for $15$ seconds
        \STATE Initiate the desired trajectory 
        \STATE Standstill for $15$ seconds
        \STATE End the MRU-P and DOT IMUs recordings
    \end{algorithmic}
    \caption{Recording Protocol For Each Experiment}\label{recording_protocol}
\end{algorithm} 

\section{Results}\label{res_section}
The following section discusses the recorded dataset, describing all recorded trajectory using both platforms and all three configurations in detail. Moreover, the length of all recorded trajectories is provided, as well as the total recorded time and the only driving time. Several selected trajectories are reconstructed from GNSS-RTK measurements. A description of the dataset structure is then provided, including the file tree structure, file naming template, and recorded measurement unit.

\subsection{Data Validation}\label{data_valid_sec}
\begin{figure}[h!]
    \centering
    \includegraphics[width = 0.95\linewidth,clip,keepaspectratio]{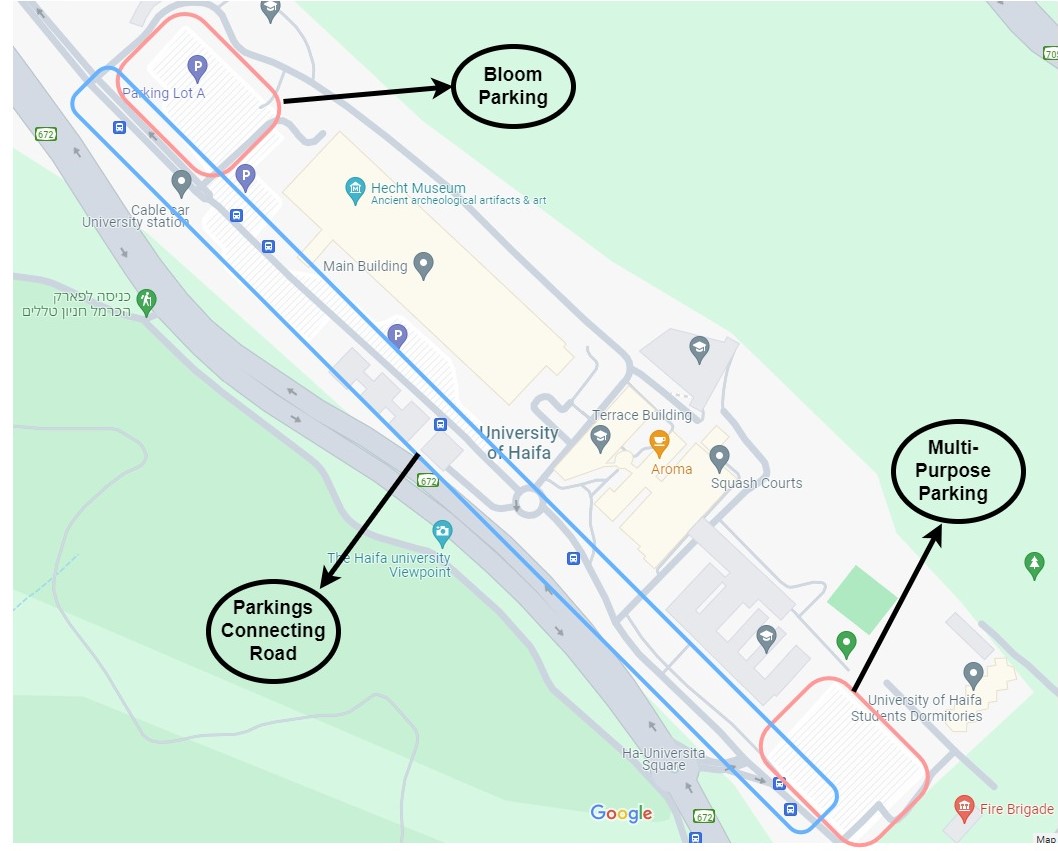}
    \caption{The University of Haifa main campus, in the top-left corner is the Bloom school of graduate studies parking lot, referred to as "Bloom", and in the bottom-right corner is the multi-purpose building parking lot, labeled as "MP". In the blue rectangle is the road connecting  between the Bloom and MP parking lots.}
    \label{univ_map}
\end{figure}

\begin{figure}[h!]
    \begin{subfigure}{.475\linewidth}
      \includegraphics[trim={0cm 0cm 0cm 2.3cm},clip,width=\linewidth]{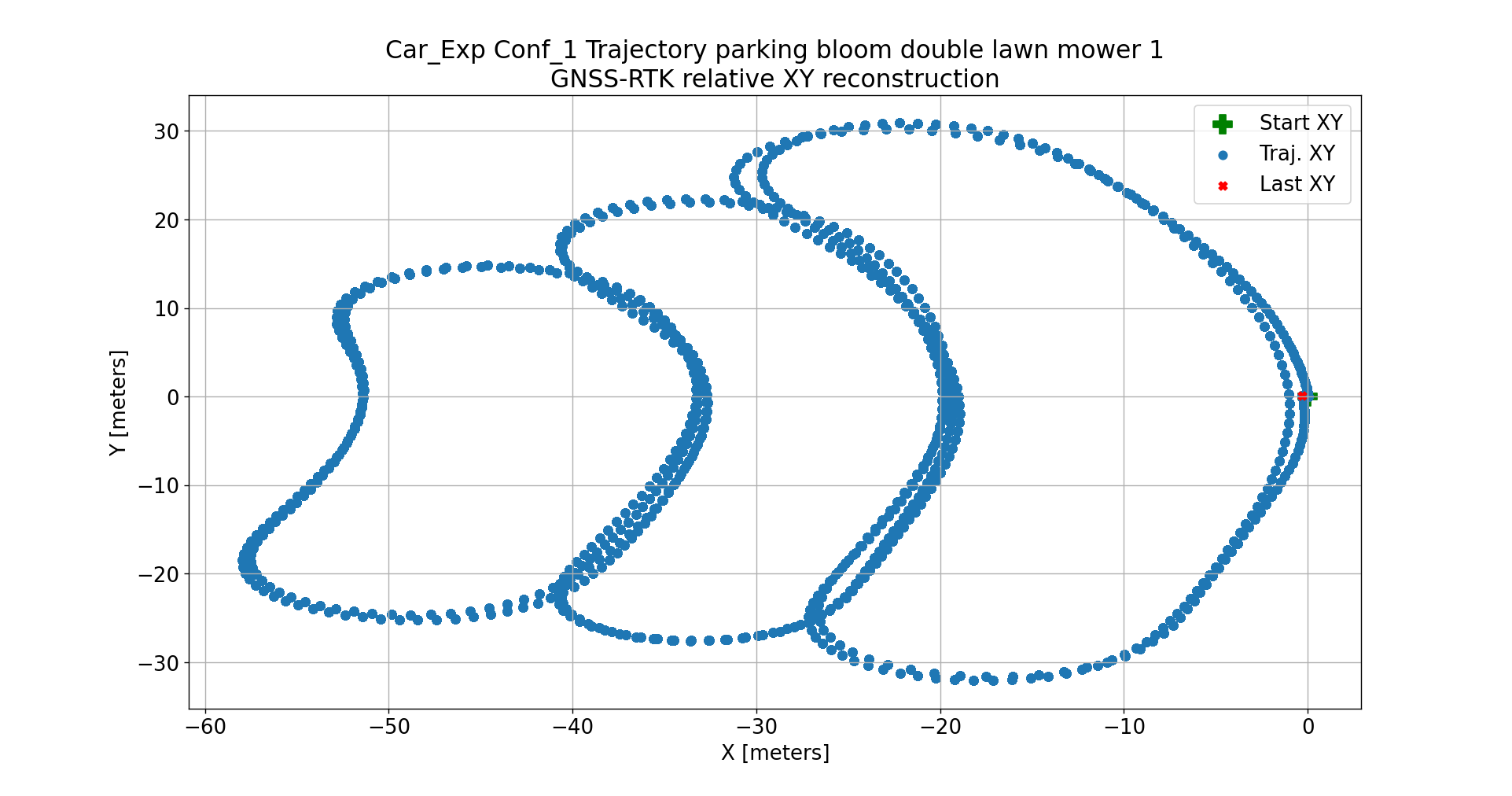}
      \caption{Bloom Double LM Traj.}
      \label{bloom_d_LM}
    \end{subfigure}\hfill 
    \begin{subfigure}{.475\linewidth}
      \includegraphics[trim={0cm 0cm 0cm 2.3cm},clip,width=\linewidth]{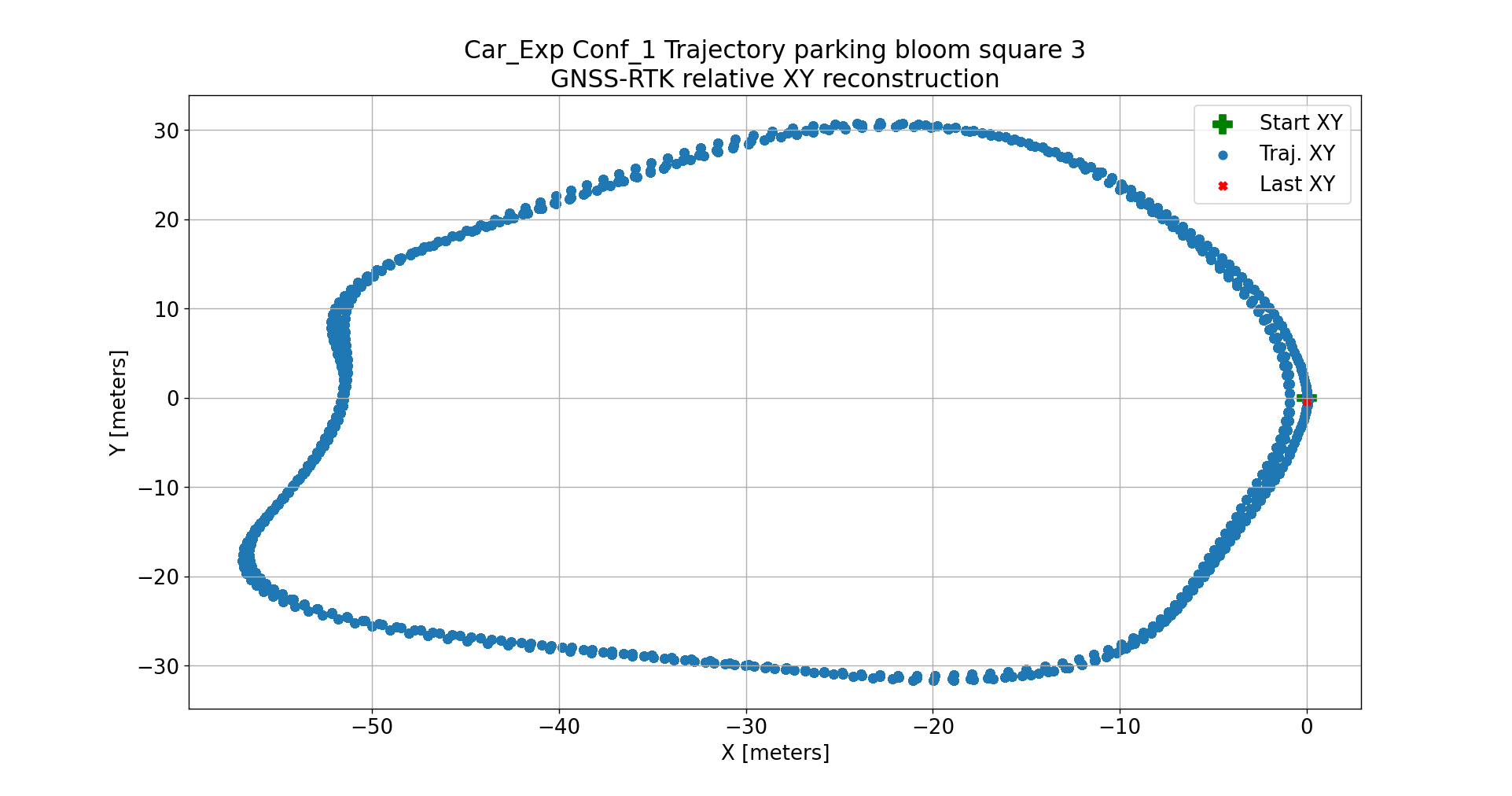}
      \caption{Bloom Square Traj.}
      \label{bloom_square_traj}
    \end{subfigure}
    \medskip 
    \begin{subfigure}{.475\linewidth}
      \includegraphics[trim={0cm 0cm 0cm 2.3cm},clip,width=\linewidth]{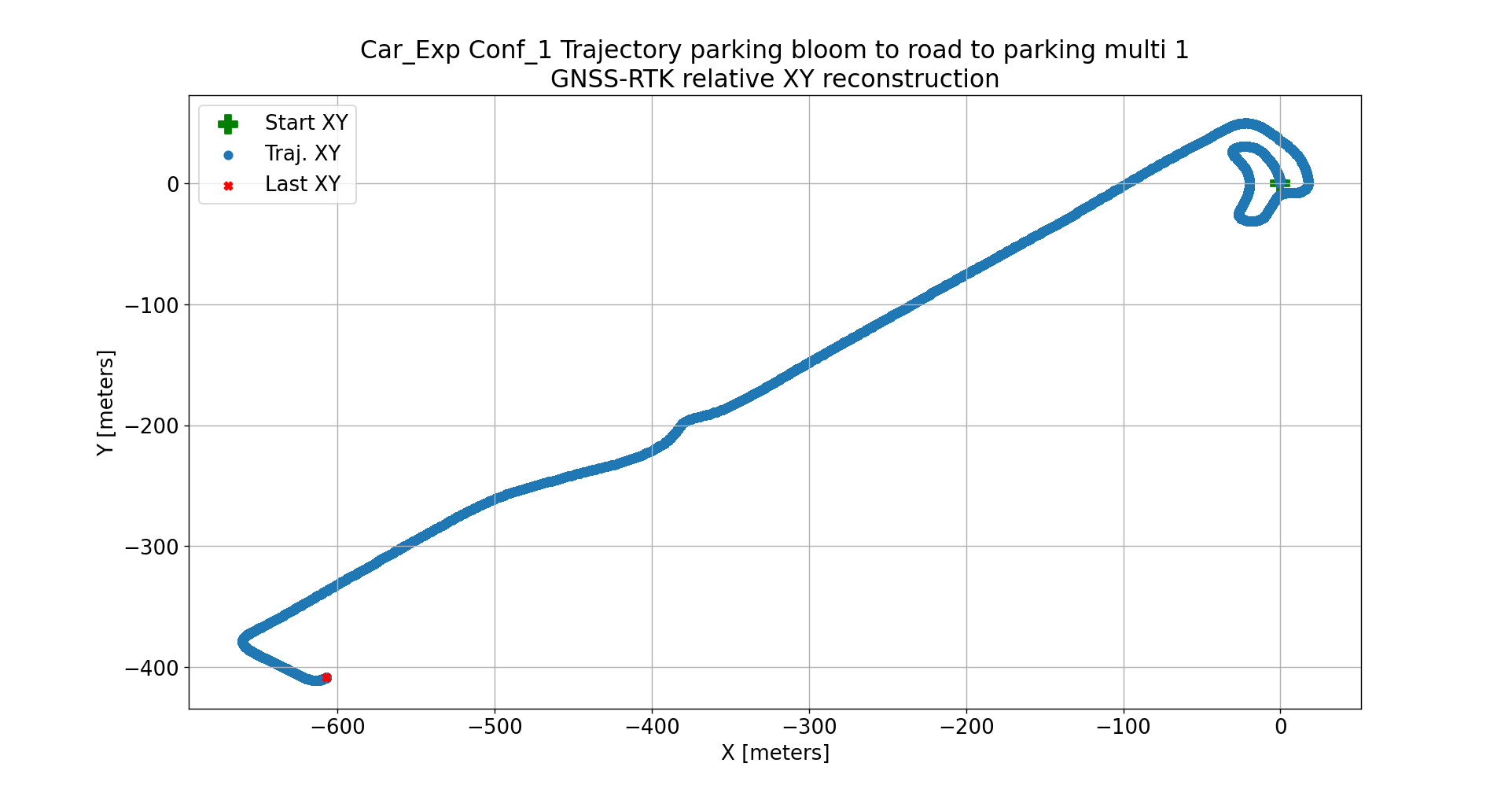}
      \caption{Bloom to MP Traj.}
      \label{bloom_to_mp_traj}
    \end{subfigure}\hfill 
    \begin{subfigure}{.475\linewidth}
      \includegraphics[trim={0cm 0cm 0cm 2.3cm},clip,width=\linewidth]{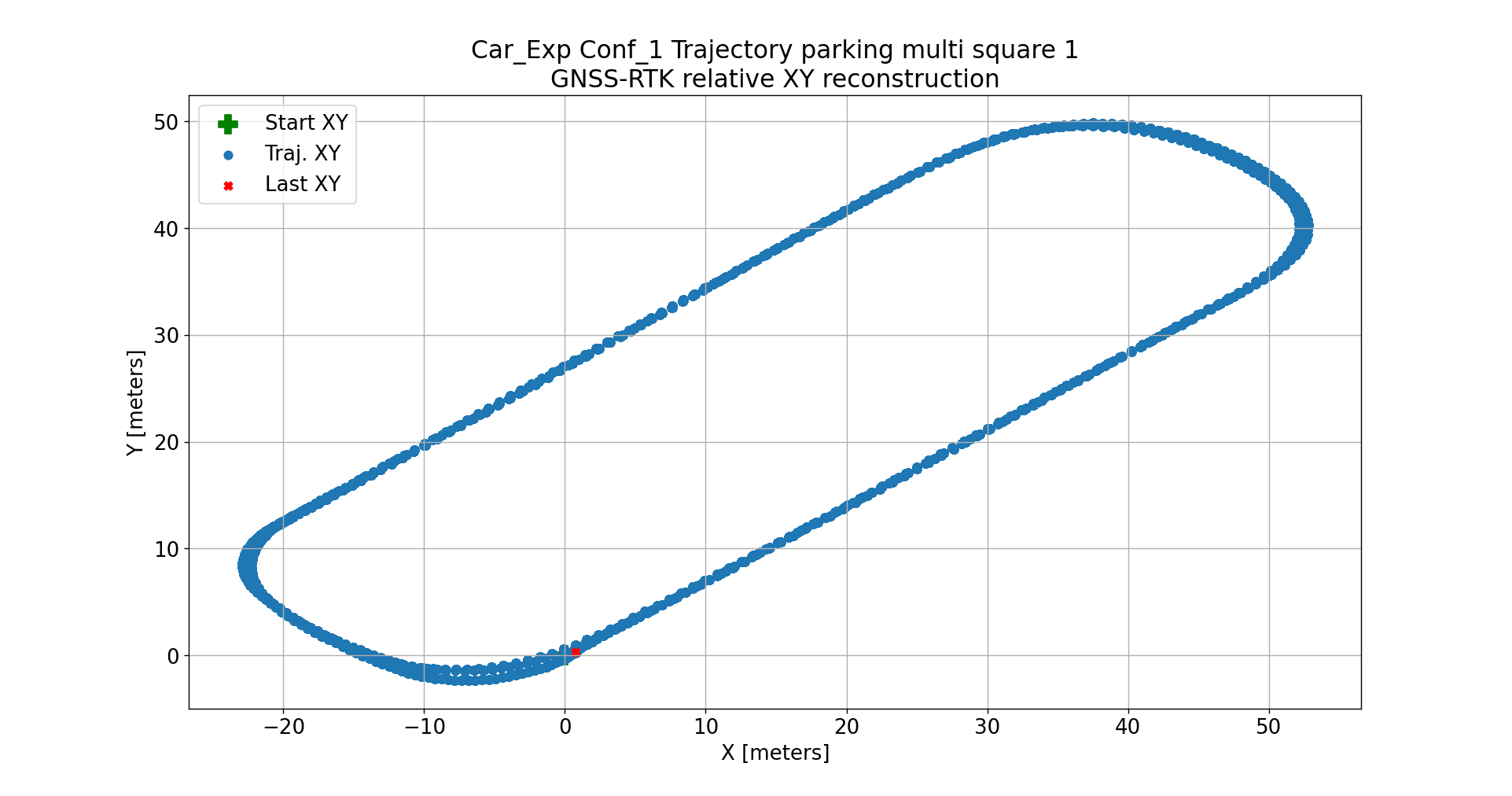}
      \caption{MP Square Traj.}
      \label{mp_square_traj}
    \end{subfigure}
\caption{Examples of four trajectories recorded with \textbf{C1} configuration mounted on top of the car.}
\label{car_gnss_plots}
\end{figure}

\begin{figure}[h!]
    \begin{subfigure}{.475\linewidth}
      \includegraphics[trim={0cm 0cm 0cm 2.3cm},clip,width=\linewidth]{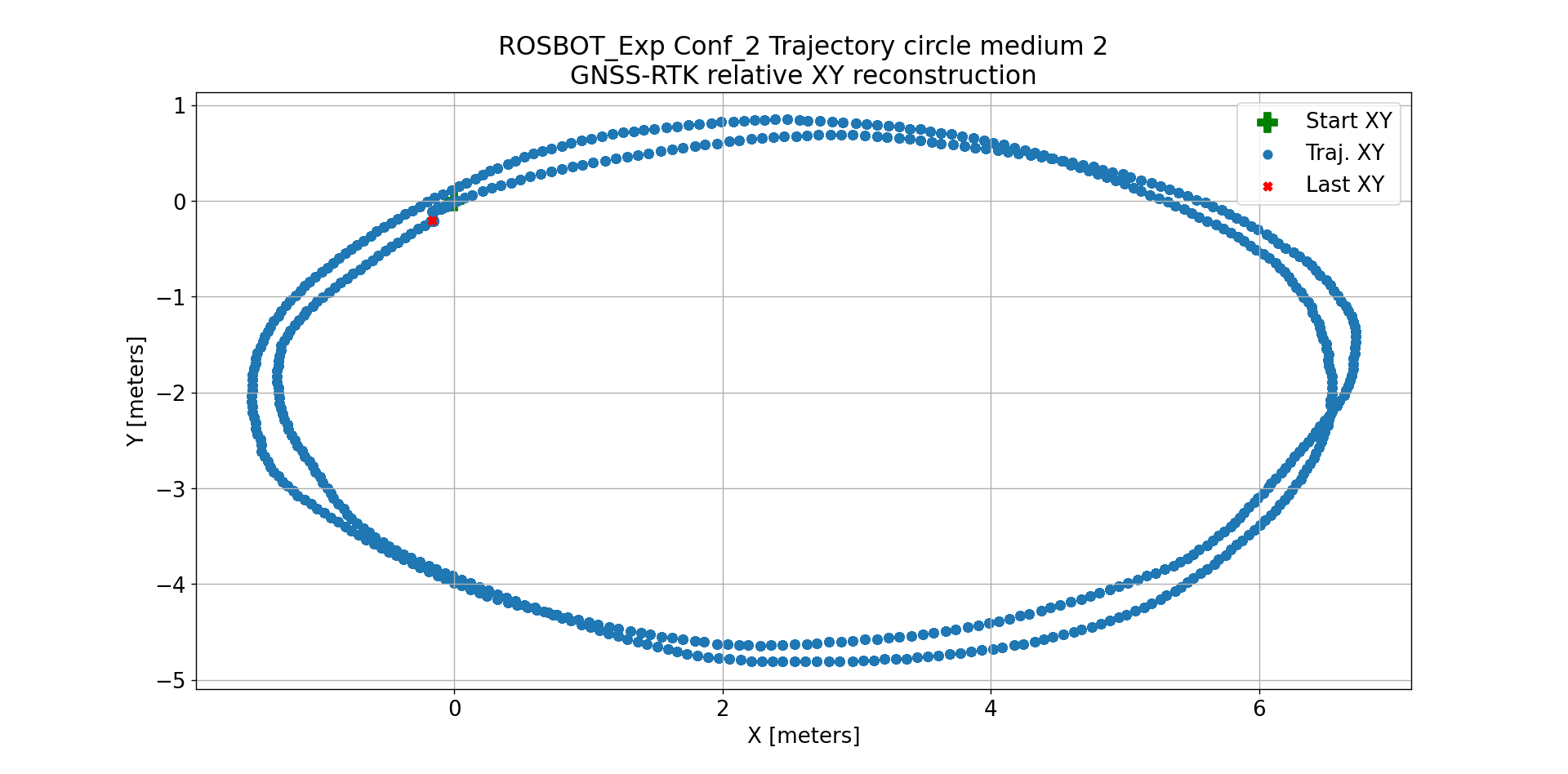}
      \caption{Circle Medium Speed}
      \label{circle_med}
    \end{subfigure}\hfill 
    \begin{subfigure}{.475\linewidth}
      \includegraphics[trim={0cm 0cm 0cm 2.3cm},clip,width=\linewidth]{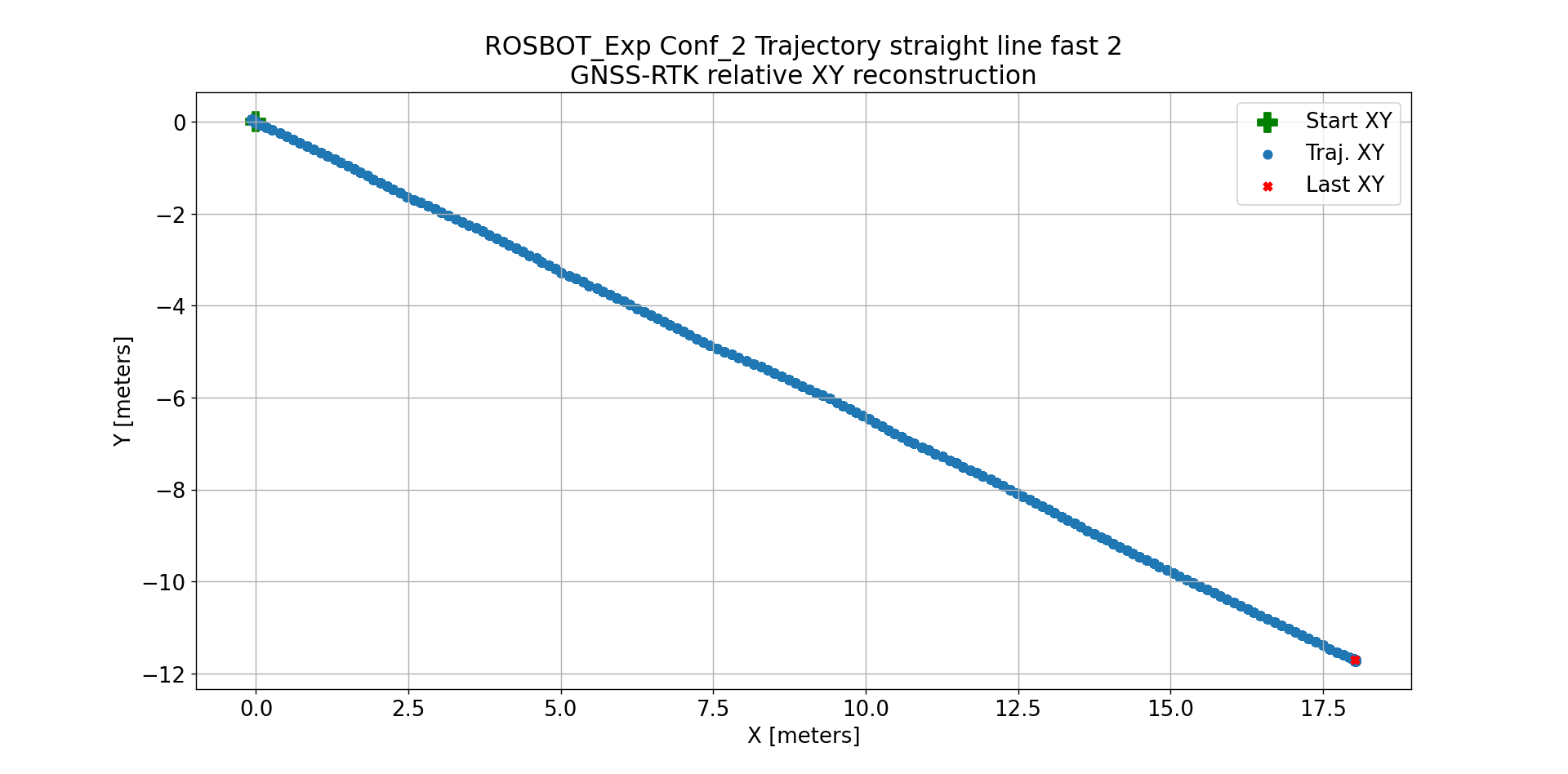}
      \caption{Line Fast Speed}
      \label{line_fast}
    \end{subfigure}
    \medskip 
    \begin{subfigure}{.475\linewidth}
      \includegraphics[trim={0cm 0cm 0cm 2.3cm},clip,width=\linewidth]{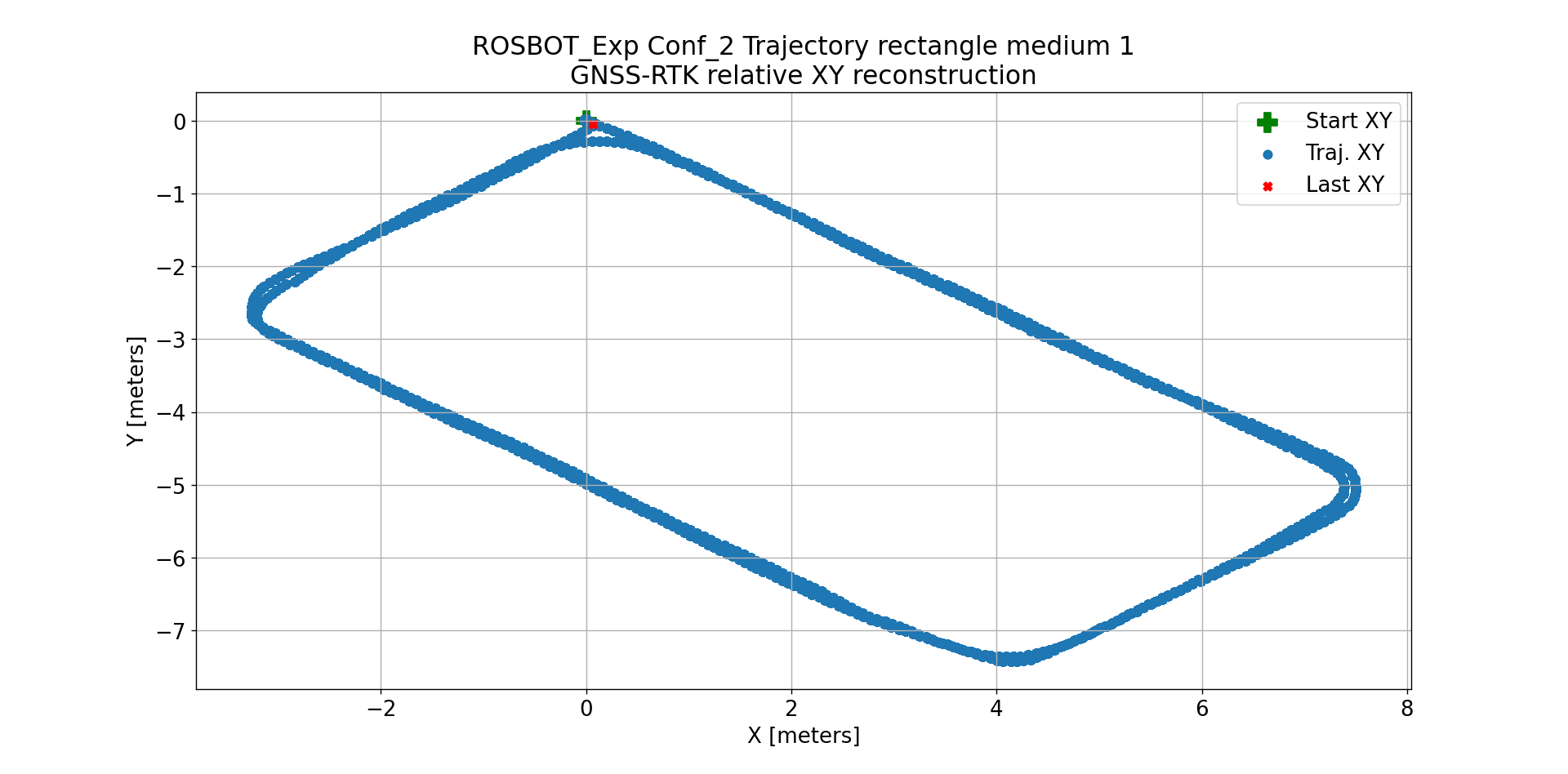}
      \caption{Rectangle Medium Speed}
      \label{rec_med}
    \end{subfigure}\hfill 
    \begin{subfigure}{.475\linewidth}
      \includegraphics[trim={0cm 0cm 0cm 2.3cm},clip,width=\linewidth]{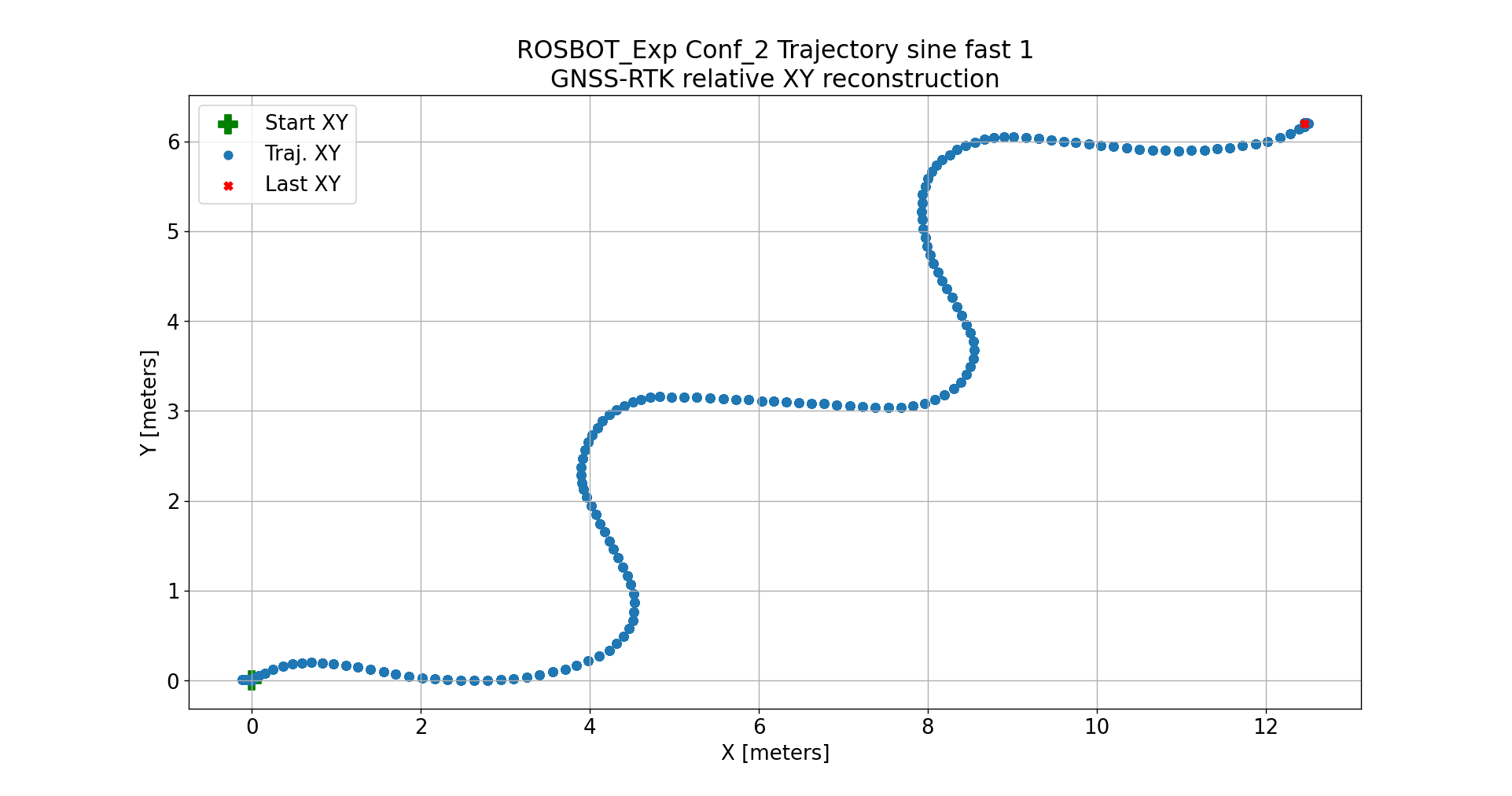}
      \caption{Sine Fast Speed}
      \label{sine_fast}
    \end{subfigure}
\caption{ Examples of four trajectories recorded with ROSbot using \textbf{C2} configuration.}
\label{rosbot_trajs}
\end{figure}

As described in Section \ref{methods_section}, two platforms of different sizes were used to record the dataset, hence different trajectories were recorded by each platform. Moreover, some variation in the number of repetitions are present in the experiment. Such variations arise from logistical reasons, such as weather, battery consumption during the experiments, and constraints related to the place where the experiments took place. \\
All trials conducted with the car took place in the University of Haifa's main campus on Mount Carmel. Most trajectories, recorded with the car, took place in two parking lots, located at opposite ends of the campus. Some trajectories involved driving back and forth between the parking lots.  Figure \ref{univ_map} shows the two parking lots and the road connecting between them. Two main patterns were recorded: a rectangle maneuver and a "lawn mower" maneuver. Additionally, a double lawn mower shape was recorded in one parking lot using only some of the sensor configurations. These patterns differ in size between the two parking lots, because one parking lot is notably larger than the other. One parking lot is adjacent to the Multi-Purpose building, and will subsequently be referred to as "MP". The other parking lot is adjacent to the Bloom School of Graduate Studies, and will subsequently be referred to as "Bloom". \\
In Table \ref{car_dynamics_and_time}, all experiments conducted with the car using \textbf{C1-C3} configurations are presented, including the type of trajectory, number of repetitions, and time description. A total of 47 trajectories were recorded with a total time of 175 minutes for each sensor. By  subtracting the stationary periods at the string and end points of the trajectories,   the driving time is 154 minutes. With nine DOTs in \textbf{C1} and eight in \textbf{C2} and \textbf{C3}, a total of 24.5 hours were recorded using the car, of which 19.6 hours were spent driving.
\begin{table}[htb!]
\centering
\caption{The trajectories recorded by the land vehicle with \textbf{C1-C3} configurations.  The total recording time is given for a single DOT IMU.}\label{car_dynamics_and_time}
\begin{adjustbox}{width=1.5\columnwidth,center,scale = 0.4}
\begin{tabular}{c|c|c|c|}
\cline{2-4}
                                              & Trajectory Type         & \begin{tabular}[c]{@{}c@{}}Number of\\ Repetitions\end{tabular} & Total Time [min]\\ \hline
\multicolumn{1}{|c|}{\multirow{7}{*}{C1}} & MP Square       & 2                                                               & 7.14       \\ \cline{2-4} 
\multicolumn{1}{|c|}{}                        & MP LM2          & 2                                                               & 7.45       \\ \cline{2-4} 
\multicolumn{1}{|c|}{}                        & MP to Bloom     & 2                                                               & 9.2        \\ \cline{2-4} 
\multicolumn{1}{|c|}{}                        & Bloom Square    & 3                                                               & 9.4        \\ \cline{2-4} 
\multicolumn{1}{|c|}{}                        & Bloom LM        & 3                                                               & 9.87       \\ \cline{2-4} 
\multicolumn{1}{|c|}{}                        & Bloom Double LM & 3                                                               & 10.63      \\ \cline{2-4} 
\multicolumn{1}{|c|}{}                        & Bloom to MP     & 2                                                               & 9.1        \\ \cline{2-4}
\multicolumn{1}{|c|}{}                        & Total  
& \textbf{17}                                                               & \textbf{62.79 }       \\ \hline
\multicolumn{1}{|c|}{\multirow{7}{*}{C2}} & MP Square       & 2                                                               & 6.55       \\ \cline{2-4} 
\multicolumn{1}{|c|}{}                        & MP LM           & 3                                                               & 11.6       \\ \cline{2-4} 
\multicolumn{1}{|c|}{}                        & MP to Bloom     & 3                                                               & 13         \\ \cline{2-4} 
\multicolumn{1}{|c|}{}                        & Bloom Square    & 2                                                               & 5.78       \\ \cline{2-4} 
\multicolumn{1}{|c|}{}                        & Bloom LM        & 2                                                               & 6.65       \\ \cline{2-4} 
\multicolumn{1}{|c|}{}                        & Bloom Double LM & 2                                                               & 6.77       \\ \cline{2-4} 
\multicolumn{1}{|c|}{}                        & Bloom to MP     & 3                                                               & 13.47      \\ 
\cline{2-4}
\multicolumn{1}{|c|}{}                        & Total    
& \textbf{17}                                                               & \textbf{63.82 }       \\ \hline
\multicolumn{1}{|l|}{\multirow{7}{*}{  C3}} & MP Square       & 2                                                               & 6.73       \\ \cline{2-4} 
\multicolumn{1}{|l|}{}                        & MP LM           & 2                                                               & 8.76       \\ \cline{2-4} 
\multicolumn{1}{|l|}{}                        & MP to Bloom     & 2                                                               & 8.97       \\ \cline{2-4} 
\multicolumn{1}{|l|}{}                        & Bloom Square    & 2                                                               & 5.76       \\ \cline{2-4} 
\multicolumn{1}{|l|}{}                        & Bloom LM        & 2                                                               & 6.02       \\ \cline{2-4} 
\multicolumn{1}{|l|}{}                        & Bloom Double LM & 1                                                               & 3.62       \\ \cline{2-4} 
\multicolumn{1}{|l|}{}                        & Bloom to MP     & 2                                                               & 9.2        \\ \cline{2-4}
\multicolumn{1}{|c|}{}                        & Total 
& \textbf{13}                                                               & \textbf{49.06 }       \\ \hline
\multicolumn{1}{|c|}{\textbf{Total}}          & -               & \textbf{47}                                                              & \textbf{175.7}     \\ \hline
\end{tabular}
\end{adjustbox}
\end{table}
\noindent The experiments conducted using the ROSbot were also carried out in the main campus, specifically in the tennis court. Several trajectories were recorded, including rectangular, circular, and sinusoidal trajectories. Additional patterns were performed with some of the configurations, such as a square and a straight line. Each movement pattern was repeated several times at different speeds categorized by slow, medium, and fast.\\
Table \ref{rosbot_dynamics_and_time} presents the experiments made with ROSbot mobile robot . The column "Speed Values" describes the number of times each trajectory was repeated at different speeds.  If "1" is shown in the "Speed Values" column, then the trajectory was recorded only at a slow speed. Similarly, if "2" is reported,  then the trajectory was recorded at slow and medium speeds. Finally, if "3" is indicated, then all three speeds were applied in the recordings. Furthermore, each trajectory was repeated several times at each speed and the number of repetitions for each speed is shown in the column "Rep. Per Speed". 
A total of 68 trajectories were recorded by ROSbot, and considering the recording time of a single sensor, the total recording time of the trajectories was 163.1 minutes.
By subtracting the stationary periods at the start and end points of the trajectories, the driving time was 132.5 minutes. Using the car, a total of 47 trajectories were recorded, totaling  175.67 minutes of recording time for a single sensor. Subtracting the standstill period at the start and end of each trajectory produces 154.52 minutes of driving time when considering a single sensor recording time.
\begin{table}[h!]
\centering
\caption{ROSbot recorded trajectories with \textbf{C1-C3} configurations. The total recording time is given for a single DOT IMU.}\label{rosbot_dynamics_and_time}
\begin{adjustbox}{width=1.5\columnwidth,center,scale = 0.5}
\begin{tabular}{l|c|c|c|c|c|}
\cline{2-6}
\multicolumn{1}{c|}{}                         & Trajectory Type   & \multicolumn{1}{c|}{\begin{tabular}[c]{@{}c@{}}Speed\\ Values\end{tabular}} & \multicolumn{1}{c|}{\begin{tabular}[c]{@{}c@{}}Rep. Per\\ Speed\end{tabular}} & \multicolumn{1}{c|}{\begin{tabular}[c]{@{}c@{}}Total\\ Repetitions\end{tabular}} & Total Time  [min]    \\ \hline
\multicolumn{1}{|c|}{\multirow{6}{*}{\textbf{C1}}} & Circle    & 3                                                                               & 2                                                                             & 6                                                                                & 12.91           \\ \cline{2-6} 
\multicolumn{1}{|c|}{}                        & Rectangle & 3                                                                               & 2                                                                             & 6                                                                                & 14.8            \\ \cline{2-6} 
\multicolumn{1}{|c|}{}                        & Square    & 3                                                                               & 2                                                                             & 6                                                                                & 11.17           \\ \cline{2-6} 
\multicolumn{1}{|c|}{}                        & Sine      & 3                                                                               & 2                                                                             & 6                                                                                & 10.37           \\ \cline{2-6} 
\multicolumn{1}{|c|}{}                        & Line      & 3                                                                               & 1                                                                             & 3                                                                                & 6.35            \\ \cline{2-6} 
\multicolumn{1}{|c|}{}                        & Total     & -                                                                              & -                                                                            & \textbf{27}                                                                      & \textbf{55.6}   \\ \hline
\multicolumn{1}{|l|}{\multirow{5}{*}{\textbf{ C2}}} & Circle    & 3                                                                               & 2                                                                             & 6                                                                                & 18.28           \\ \cline{2-6} 
\multicolumn{1}{|l|}{}                        & Rectangle & 3                                                                               & 2                                                                             & 6                                                                                & 20.11           \\ \cline{2-6} 
\multicolumn{1}{|l|}{}                        & Sine      & 3                                                                               & 2                                                                             & 6                                                                                & 11.41           \\ \cline{2-6} 
\multicolumn{1}{|l|}{}                        & Line      & 2                                                                               & 2                                                                             & 4                                                                                & 6.66            \\ \cline{2-6} 
\multicolumn{1}{|l|}{}                        & Total     & -                                                                              & -                                                                            & \textbf{22}                                                                      & \textbf{56.46}  \\ \hline
\multicolumn{1}{|l|}{\multirow{5}{*}{\textbf{ C3}}} & Circle    & 3                                                                               & 2                                                                             & 6                                                                                & 17.28           \\ \cline{2-6} 
\multicolumn{1}{|l|}{}                        & Rectangle & 3                                                                               & 2                                                                             & 6                                                                                & 20.41           \\ \cline{2-6} 
\multicolumn{1}{|l|}{}                        & Sine      & 3                                                                               & 2                                                                             & 6                                                                                & 11.61           \\ \cline{2-6} 
\multicolumn{1}{|l|}{}                        & Line      & 1                                                                               & 1                                                                             & 1                                                                                & 1.75            \\ \cline{2-6} 
\multicolumn{1}{|l|}{}                        & Total     & -                                                                              & -                                                                            & \textbf{19}                                                                      & \textbf{51.05}  \\ \hline
\multicolumn{1}{|c|}{\textbf{Total}}          & -        & -                                                                              & -                                                                            & \textbf{68}                                                                      & \textbf{163.11} \\ \hline
\end{tabular}
\end{adjustbox}
\end{table}
\begin{figure}[h!]
\centering
        \includegraphics[width=0.8\linewidth, keepaspectratio]{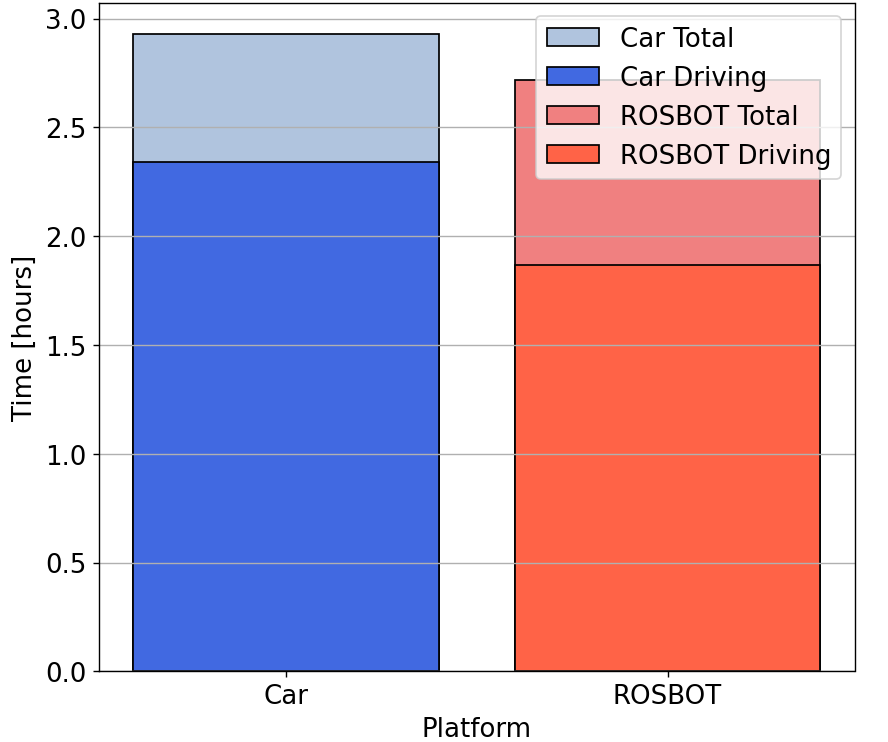}
          \caption{Recording time for a single DOT showing the total recording time (stationary and driving) and driving time.}
          \label{platform_time_comp_fig}
\end{figure}
\begin{figure}[h!]
    \centering
    \begin{adjustbox}{minipage=\linewidth, scale=1}
        \includegraphics[width=1.0\linewidth, keepaspectratio]{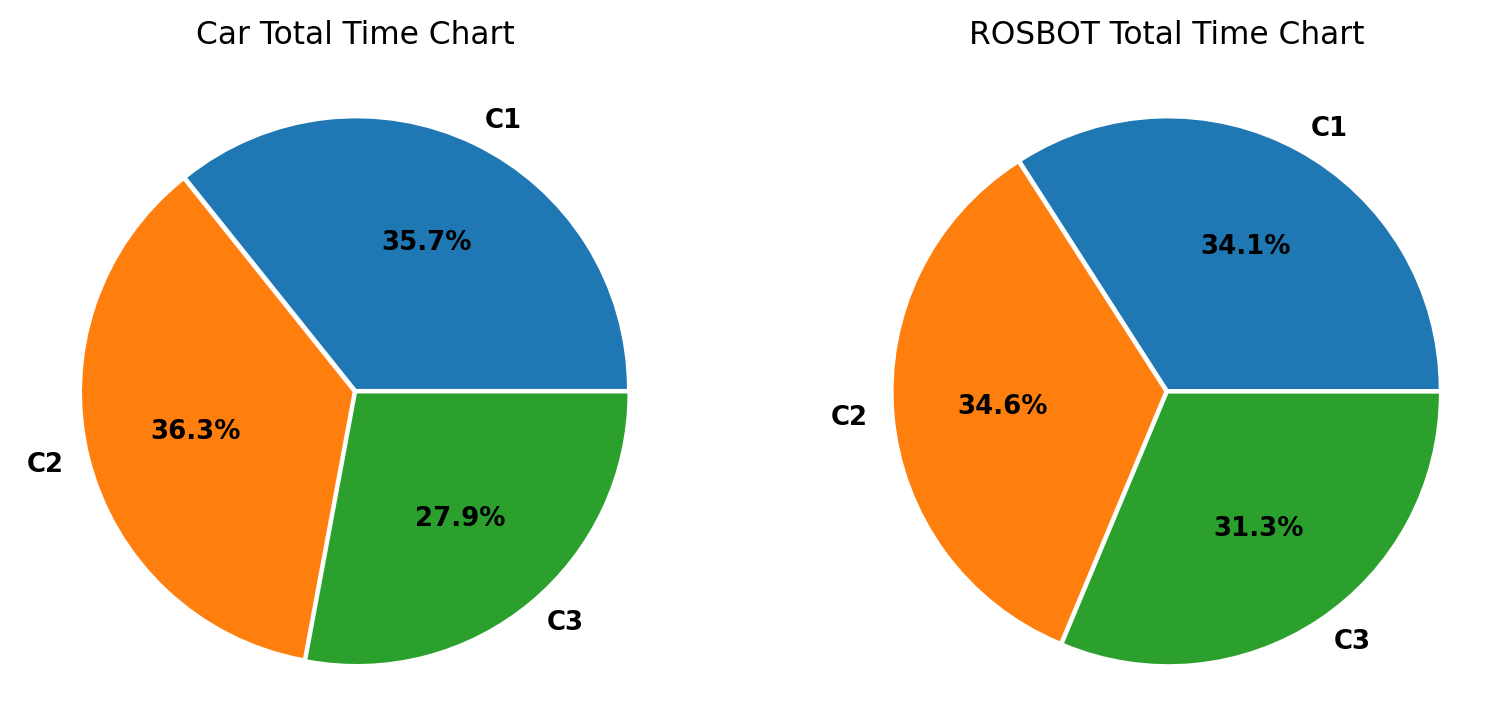}
          \caption{The relative percentage of recording time for each configuration on both platforms.}
          \label{time_conf_per_platform_pie}
    \end{adjustbox}
\end{figure}
All DOTs, in all three configurations, have a total recording time of $47.14$ hours. Subtracting the stationary periods at the start and end of each trajectory gives a total driving time of $35$ hours. 
Figure \ref{platform_time_comp_fig} summarizes the total recording time versus driving only for a single DOT sensor on both platforms. From a different point of view, Figure \ref{time_conf_per_platform_pie} gives two pie charts showing the percentage of each configuration time on all recordings, both for the car and the mobile robot.\\
\noindent Next, we randomly selected four trajectories form each platform for visualization purposes. Figure~\ref{car_gnss_plots} illustrates the trajectories   using the GNSS-RTK measurements for \textbf{C1} configuration mounted on top of the car. The selected paths are Bloom double LM, Bloom Square, Bloom parking to MP parking, and MP square. When comparing the square maneuver in the two parking lots, "Bloom Square" in Figure~\ref{bloom_square_traj} and "MP Square" in Figure~\ref{mp_square_traj}, two things are evident. First, the dimensions of the parking lots differ significantly. Secondly, both square maneuvers do not actually resemble geometric squares, due to the fact that each square maneuver was performed along the perimeter of the parking lot, following the actual shape of the parking lot. Figure \ref{rosbot_trajs} shows the four trajectories recorded by the ROSbot with \textbf{C2} configuration. Among the trajectories selected are: circle trajectory at medium speed, straight line trajectory at fast speed, rectangle trajectory at medium speed, and a sine trajectory at fast speed. Notice that the circle and rectangle trajectories were repeated several times during the same recording.


\subsection{Dataset Structure}\label{data_Recs}
\begin{figure}[h!]
\centering
        \includegraphics[width=0.9\linewidth, clip, keepaspectratio]{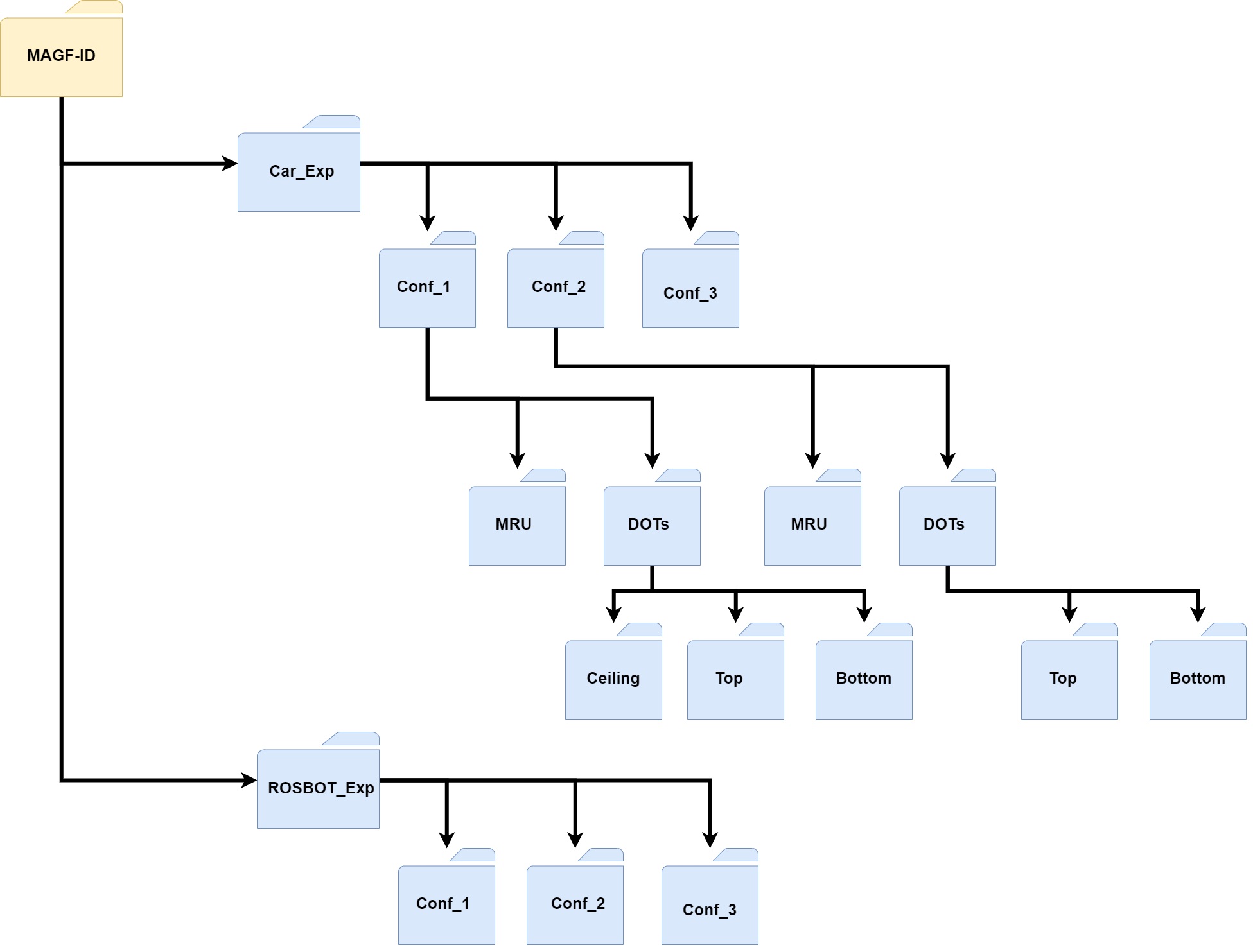}
\caption{MAGF-ID folder tree as stored in the project GitHub}\label{datase_folder_tree}
\end{figure}

This section, describes the organization of the recordings in the dataset including the hierarchy of the folders and subfolders.  The dataset is organized in a hierarchical manner with the first level of folders representing the platforms. "Car\_Exp" and "ROSbot\_Exp" refer to the car and ROSbot platforms, respectively. There are three subfolders in each platform's folder, named "Conf\_1", "Conf\_2", and "Conf\_3", which correspond to \textbf{C1-C3} configurations. Each of these folders contains the recordings as presented in Tables \ref{car_dynamics_and_time} - \ref{rosbot_dynamics_and_time}. Each configuration folder contains two subfolders, one for each sensor, "MRU" for the MRU-P, and "DOTs" for the DOT sensors. The MRU folder contains the MRU-P inertial sensors and GNSS-RTK recordings, which are saved in the form of comma-separated value files, or ".csv". The DOTs folder contains their inertial readings in two to three subfolders. For experiments conducted using \textbf{C1}, there are three folders named: "Bottom", "Top", and "Ceiling", each representing a level at which the nine DOTs are located. For experiments recorded using \textbf{C2} and \textbf{C3}, there are only two folders, "Top" and "Bottom". In the same manner as the MRU files, the DOT recordings are also stored in the format of a CSV file. Figure \ref{datase_folder_tree} illustrates the folder structure of the dataset.\\
%
\noindent 
All recordings from the MRU-P have the following structure: "MRU measurs \verb|<|trajectory name\verb|>| \verb|<|speed and repetition number\verb|>|.csv". For example, the trajectory labeled double LM recorded with the car in Bloom parking lot is named "MRU measurs Bloom double LM 1.csv". The parent folders indicate the configuration and platform used. In a similar manner, recordings of DOTs will be named by: "\verb|<|DOT number\verb|>| \verb|<|trajectory name\verb|>| \verb|<|speed and repetition number\verb|>|.csv". As an example, the rectangle recorded using the ROSbot platform driving at slow speed will be named as "DOT 10 rectangle slow 1.csv". Notice, that in the "Top" and "Bottom" folders, four files will appear, one for each of the four DOTs in the upper or lower levels of the configuration. The "Ceiling" folder, however, will only contain one file per trajectory. The parent folders indicate the platform and the configuration used, as before.\\
\noindent The MRU-P CSV files contain the accelerometer, gyroscope, magnetometer, and GNSS-RTK  measurements. Table \ref{mru_files_headers} shows the header descriptions of the MRU-P CSV files, as well as the units of measurement. The DOT IMUs CSV files contain the internal DOTs clock readings, as well as the accelerometer and gyroscope measurements. Table \ref{dots_header_desc} gives the DOTs CSV file header names, units of measurement, and the header descriptions.

\begin{table}[h!]
\centering
\caption{MRU-P ".csv" file descriptions.}\label{mru_files_headers}
\begin{tabular}{|c|c|c|}
\hline
\begin{tabular}[c]{@{}c@{}} Header\\ Name\end{tabular}
                                & \begin{tabular}[c]{@{}c@{}}Output\\ Measurement Units\end{tabular} & \begin{tabular}[c]{@{}c@{}}Header\\ Description\end{tabular}                                                   \\ \hline
\multicolumn{1}{|c|}{Lat\_GNSS}  & $deg$                                                        & \multirow{3}{*}{\begin{tabular}[c]{@{}c@{}}Position Vector: Latitude \\ Longitude and Height\end{tabular}} \\ \cline{1-2}
\multicolumn{1}{|c|}{Long\_GNSS} & $deg$                                                        &                                                                                                                \\ \cline{1-2}
\multicolumn{1}{|c|}{Height\_GNSS}     & $meter$                                                          &                                                                                                                \\ \hline
\multicolumn{1}{|c|}{Acc\_X}   & $g$                                                        & \multirow{3}{*}{\begin{tabular}[c]{@{}c@{}}Accelerometer\\ Measurements\\ \end{tabular}} \\ \cline{1-2}
\multicolumn{1}{|c|}{Acc\_Y}  & $g$                                                         &                                                                                                                \\ \cline{1-2}
\multicolumn{1}{|c|}{Acc\_Z}     & $g$                                                          &                                                                                                                \\ \hline
\multicolumn{1}{|c|}{Gyro\_X}   & $\degree /s$                                                     & \multirow{3}{*}{Gyroscopes Measurements}                                                                      \\ \cline{1-2}
\multicolumn{1}{|c|}{Gyro\_Y}     & $\degree /s$                                                         &                                                                                                                \\ \cline{1-2}
\multicolumn{1}{|c|}{Gyro\_Z}      & $\degree /s$                                                         &                                                                                                                \\ \hline
\multicolumn{1}{|c|}{Magn\_X}   & $\mathrm {n} T$                                                     & \multirow{3}{*}{Magnetometer Measurements}                                                                      \\ \cline{1-2}
\multicolumn{1}{|c|}{Magn\_Y}     & $\mathrm {n} T$                                                         &                                                                                                                \\ \cline{1-2}
\multicolumn{1}{|c|}{Magn\_Z}      & $\mathrm {n} T$                                                         &                                                                                                                \\ \hline
\end{tabular}
%
\end{table}
\begin{table}[h!]
\centering
\caption{DOT ".csv" file descriptions.}\label{dots_header_desc}
\begin{tabular}{|c|c|c|}
\hline
\begin{tabular}[c]{@{}c@{}}Header\\ Name\end{tabular} & \begin{tabular}[c]{@{}c@{}}Units of\\ Measurement\end{tabular} & \begin{tabular}[c]{@{}c@{}}Header\\ Description\end{tabular}                                 \\ \hline
sampleTimeFine                                        & $ms$                                                             & \begin{tabular}[c]{@{}c@{}}Time in\\ mili-seconds  \\ \end{tabular} \\ \hline
Acc\_X                                                & $m / {s^2}$                                                            & \multirow{3}{*}{\begin{tabular}[c]{@{}c@{}} Accelerometer\\ Measurements\end{tabular}}    \\ \cline{1-2}
Acc\_Y                                                & $m / {s^2}$                                                            &                                                                                              \\ \cline{1-2}
Acc\_Z                                                & $m / {s^2}$                                                             &                                                                                              \\ \hline
Gyr\_X                                                & $ \degree / s$                                                            & \multirow{3}{*}{\begin{tabular}[c]{@{}c@{}} Gyroscope\\ Measurements\end{tabular}}        \\ \cline{1-2}
Gyr\_Y                                                & $\degree / s$                                                            &                                                                                              \\ \cline{1-2}
Gyr\_Z                                                & $\degree / s$                                                            &                                                                                              \\ \hline
\end{tabular}
\end{table}

\section{Conclusion}\label{conc_sec}
Inertial sensors are required in various fields such as vehicles, robotics, healthcare, and the Internet of Things (IoT). They are also mounted on different platforms such as vehicles, drones, and smartwatches. As a result, there has been a continuous increase in inertial sensing related topics focusing on improving inertial sensors' accuracy, efficiency, and configuration. Despite the need, there are no available datasets for GFINS and MIMU configurations. To fill this gap and stimulate further research in this field, we used eight to nine IMUs to construct GFINS and MIMU datasets. The sensors were arranged in three different sensor configurations and mounted on a land vehicle and a mobile robot. These sensors can be used to define and evaluate different types of MIMU and GFINS architectures. In all configurations, there was a GNSS-RTK to provide ground truth trajectories.  While the first configuration had nine IMUs, the second and third configurations had eight. All sensors were positioned in the same orientation and direction on various structures. In total, the MAGF-ID contains $115$ trajectories with a total of $35$ hours of inertial data and associated ground truth trajectories. The recordings followed a protocol which involved starting the sensors manually at the same time, then shaking the platform for synchronization purposes. All of the recorded data is accessible through our GitHub repository \href{https://github.com/ansfl/MAGF-ID}{MAGF-ID GitHub}.

\section{Acknowledgements}
Z.Y, Y.S, I.S. N.P.H, S.M and D.S are all supported by the Maurice Hatter Foundation.
Z.Y. is also supported by the University of Haifa presidential scholarship for outstanding students on a direct Ph.D. track.



\bibliographystyle{IEEEtran}
\bibliography{bio.bib}

\end{document}